\documentclass[conference]{IEEEtran}
\usepackage{breakurl}
\usepackage{xurl}
\usepackage{subcaption}
\usepackage{tikz}
\usepackage{hyperref}
\usepackage{pgfplots}
\usepackage{array,multirow,graphicx}
\usetikzlibrary{patterns}
\usepackage{rotating}
\usepackage{amsmath}

\usepackage{algorithm, algpseudocode}
\usepackage{xcolor}
\usepackage{placeins}
\usetikzlibrary{decorations.pathreplacing,positioning}
\usepackage{graphicx} 
\usepackage{tabularx} 
\usepackage[noadjust]{cite}
\DeclareRobustCommand*{\IEEEauthorrefmark}[1]{%
  \raisebox{0pt}[0pt][0pt]{\textsuperscript{\footnotesize\ensuremath{#1}}}}
  \usepackage[misc]{ifsym}
\newcommand{\corrAuthor}{$^{\textrm{\Letter}}$}

\begin{document}
\bstctlcite{IEEEexample:BSTcontrol}
\title{Recovery of Distributed Iterative Solvers for \\ Linear Systems Using Non-Volatile RAM}


\author{\authorblockN{Yehonatan Fridman\authorrefmark{1,2},
Yaniv Snir\authorrefmark{1,3},
Harel Levin\authorrefmark{4,5},
Danny Hendler\authorrefmark{1},
Hagit Attiya\authorrefmark{6} and
Gal Oren\authorrefmark{5,6} \corrAuthor}\\
\authorblockA{\authorrefmark{1}Department of Computer Science, Ben-Gurion University of the Negev}
\authorblockA{\authorrefmark{2}Israel Atomic Energy Commission}
\authorblockA{\authorrefmark{3}Google}
\authorblockA{\authorrefmark{4}Mobileye Vision Technologies}
\authorblockA{\authorrefmark{5}Scientific Computing Center, Nuclear Research Center -- Negev}
\authorblockA{\authorrefmark{6}Department of Computer Science, Technion -- Israel Institute of Technology}\\
{\tt\small \{fridyeh, yanivsn\}@post.bgu.ac.il, harellevin@nrcn.org.il,}\\
{\tt\small hendlerd@cs.bgu.ac.il, \{hagit, galoren\}@cs.technion.ac.il}}

\maketitle

\begin{abstract}
HPC systems are a critical resource for scientific research and advanced industries. 
The increased demand for computational power and memory ushers in the \emph{exascale era}, 
in which supercomputers are designed to provide enormous computing 
power to meet these needs. 
These complex supercomputers consist of numerous compute nodes and are consequently 
expected to experience frequent faults and crashes.

Mathematical solvers, in particular, \emph{iterative linear solvers} are 
key building block in numerous large-scale scientific applications. 
Consequently, supporting the recovery of \emph{distributed} solvers is necessary for 
scaling scientific applications to exascale platforms. 
Previous recovery methods for iterative solvers are based on Checkpoint-Restart (CR),
which incurs high fault tolerance overhead, or intrinsic fault tolerance,
which require extra computation time to converge after failures. 

\emph{Exact state reconstruction} (\emph{ESR}) was proposed as an alternative mechanism 
to alleviate the impact of frequent failures on long-term computations. 
ESR has been shown to provide exact reconstruction of the computation state while 
avoiding the need for costly checkpointing. 
However, ESR currently relies on volatile memory for fault tolerance, 
and must therefore maintain redundancies in the RAM of multiple nodes. 
This not only incurs high memory overhead but also prevents ESR from being 
\emph{fully resilient}, that is, resilient against a full system crash. 

Recent supercomputer designs feature emerging \emph{non-volatile RAM} (\emph{NVRAM}) technology, 
for example, the exascale \emph{Aurora} that is planned to consist of Intel Optane™ DCPMM. 
This paper investigates how NVRAM can be utilized to devise an enhanced ESR-based recovery 
mechanism that is more efficient and provides full resilience. 
Our mechanism, called \emph{in-NVRAM} ESR, provides full resiliency while 
significantly reducing both the memory footprint and the time overhead 
in comparison with the original ESR design (\emph{in-RAM} ESR). 
In-NVRAM ESR is based on a novel MPI One-Sided Communication (OSC) over RDMA implementation, 
which was optimized and applied for using NVRAM to store recovery data for iterative linear solvers.

The source code used in this work, as well as the benchmarks and other relevant sources, are available at: \textcolor{blue}{\url{https://github.com/Scientific-Computing-Lab-NRCN/In-NVRAM-ESR.git}}.
\end{abstract}
\maketitle

\section{Introduction}
\subsection{Fault Tolerance in Supercomputing}
The past decade has seen a skyrocketing increase in the demand for high-performance computing 
(HPC) systems, in order to meet the computing power needs of resource-hungry applications 
in various science domains.
This ushered in a new exascale era of stronger and more complex supercomputers.
For example, the first exascale supercomputer, \textit{Frontier} \cite{schneider2022exascale}, 
consists of more than 8 million compute cores that provide peak computing power of 1.102 Exaflop/s. 
With the increase in complexity and the number of compute nodes, comes increased vulnerability to failures.
Already for earlier generations of petascale supercomputers, 
Schroeder and Gibson~\cite{SchroederGibson2007} showed that in certain situations, 
applications are forced into recovery more than twice a day. 
It is further anticipated that exascale systems will experience various kinds of faults 
every few hours or even every few minutes, in spite of hardware-based protection 
mechanisms~\cite{cappello2009toward,canal2020predictive}, 
leading to a growing need for mechanisms to ensure efficient recovery from failures.

\emph{Mathematical solvers} are a key component of scientific applications, especially in HPC, 
accounting for 50-90$\%$ of their operations~\cite{heroux2019recent}. 
Efforts were made to develop and maintain distributed, 
efficient and scalable libraries for these solvers. 
A prominent example is \emph{Trilinos}~\cite{heroux2005overview}, 
a collection of distributed and parallel reusable scientific software libraries
that is widely used in high-performance scientific computing and includes linear, 
non-linear, transient and optimization solvers.
Supporting the recovery of mathematical solvers is therefore necessary for scaling scientific 
applications and performing them on exascale platforms. 
Indeed, prior research~\cite{heroux2009software,chien2017final,chien2017exploring, dun2015flexible,bridges2012fault}
studied soft and hard faults resilience aspects in Trilinos.

Checkpoint-Restart (CR) techniques using non-volatile storage such as HDDs and SSDs 
are the most general and direct mechanisms to support recovery of scientific 
applications~\cite{moody2010design, ansel2009dmtcp}. 
Extensive work was done to provide improved CR models in order to minimize 
performance loss~\cite{naksinehaboon2008reliability, liu2008optimal}. 
However, CR has inherent limitations, especially for scientific computations with 
large data sets, where checkpointing the partial image of an application 
(let alone the full one) requires to transfer large amounts of data to high-latency devices. 
Excessive checkpointing will result in total performance degradation.   

Other work investigates fault-tolerance aspects in mathematical solvers, 
exploiting intrinsic attributes of solvers to tolerate faults, 
alleviating the need for checkpointing~\cite{heroux2014toward,bridges2012fault,bosilca2004recovery,sao2013self}. 
For example, Sao and Vuduc \cite{sao2013self} investigate self-stabilizing iterative
solvers that require only lightweight tests for fault-detection. 
However, these solutions typically do not guarantee correct recovery, 
or require a significant extra computation time to converge after recovery.

\subsection{NVRAM in HPC}
Next-generation supercomputers are expected to incorporate novel 
\textit{non-volatile memory} (\textit{NVM}); 
for example, the flagship \textit{Aurora}~\cite{Argonne1}
will integrate NVM devices such as the Intel Optane™ DCPMM. 
When configured in App-Direct mode, these devices are byte-addressable and 
can be used by processes as \emph{non-volatile random access memory} 
(\emph{NVRAM}).\footnote{Alternatively, they may be configured in Memory or Flat mode, in which case that provide an extension to 
the volatile memory pool of the application \cite{patil2021nvm,peng2019system}.}
NVRAM's ability to retain data even after node failures opens new possibilities for HPC,
most notably, as a medium for data persistence upon node or process failure and recovery. 

Prior work on the use of NVRAM in HPC environments was confined to three main direct use-cases: 
(1) memory expansion to enable larger memory scientific workloads~\cite{fridman2021assessing,patil2019performance,weiland2019early,patil2021nvm},
(2) fast storage for diagnostics~\cite{fridman2021assessing,hennecke2020daos,weiland2018exploiting}, and 
(3) fast persistence area for checkpointing~\cite{fridman2021assessing,ren2019easycrash}.
Use cases (2) and (3) rely on the non-volatility of NVRAM for fast storage as a substitute for standard storage mediums~\cite{weiland2018exploiting,hennecke2020daos,fridman2021assessing}. 
For example, the DAOS storage server~\cite{hennecke2020daos} is one of the promising storage systems 
for massively distributed NVM (NVMeSSD/+NVRAM);
it is already in use by supercomputers at the top of IO500 list~\cite{kunkel2016establishing}. 


In contrast to these use-cases of NVRAM (for diagnostics, checkpointing and memory expansion), 
recent research investigates the recovery of concurrent data objects using NVRAM~\cite{attiya2018nesting,aksenov2021execution,golab2017recoverable,friedman2017brief,attiya2020tracking,coburn2011nv,venkataraman2011consistent}. These works focus on achieving correct recovery with marginal footprint of the recovery data and minimal time overhead mechanisms. NVRAM has limited endurance and can tolerate a limited number of writes~\cite{qureshi2009enhancing} --- a fact that increases the need for clever recovery models that emphasize on minimizing memory footprint of the persisted data. 
However, all of those works are mostly theoretical and were not geared towards HPC applications. 

\subsection{Exact State Reconstruction of Linear Iterative Solvers}
Iterative linear solvers, commonly used in HPC, offer inherent recoverability capabilities. 
They are mostly suitable for linear problems involving many variables (sometimes on the order of millions), where direct methods would be prohibitively expensive~\cite{pommerell1994memory}. 
In Trilinos, these solvers are provided in the \textit{Anasazi}~\cite{baker2009anasazi} and \textit{Belos}~\cite{bavier2012amesos2} libraries,
and include CG~\cite{nazareth2009conjugate}, GMRES~\cite{saad1986gmres}, Jacobi~\cite{rutishauser1966jacobi}, SOR~\cite{hadjidimos2000successive} and more. Given initial approximations, such methods converge to the correct result by improving the quality of their  result over time. 
This provides intrinsic tolerance to errors, as the solver has the potential to converge from any initial guess. 

The intrinsic fault tolerance to errors of iterative linear solvers introduces a trade-off with CR mechanisms: Iterative solvers can inherently tolerate inconsistent data during convergence, although this might require extra computation time to converge~\cite{ren2019easycrash,agullo2013towards}. 
In contrast, CR mechanisms checkpoint the exact state of the solver and computation continues in recovery from the last checkpoint. Calculations performed after the last checkpoint are lost upon a failure, hence frequent checkpointing minimizes the rollback cost.
However, excessive checkpointing creates significant storage access overheads and might double or even triple the memory footprint of the application.
This is a major downside for scientific simulations with large data sets~\cite{ren2019easycrash}. 
The gap between these two recovery mechanisms and their downsides calls for recovery methods that reduce the memory overhead without inaccuracies or imperfections, while crucial data is persisted directly into non-volatile yet fast memory, for seamless fault tolerance.

\begin{figure*}
\begin{subfigure}{0.25\textwidth}
  \centering
  \includegraphics[width=4.2cm,height=3cm]{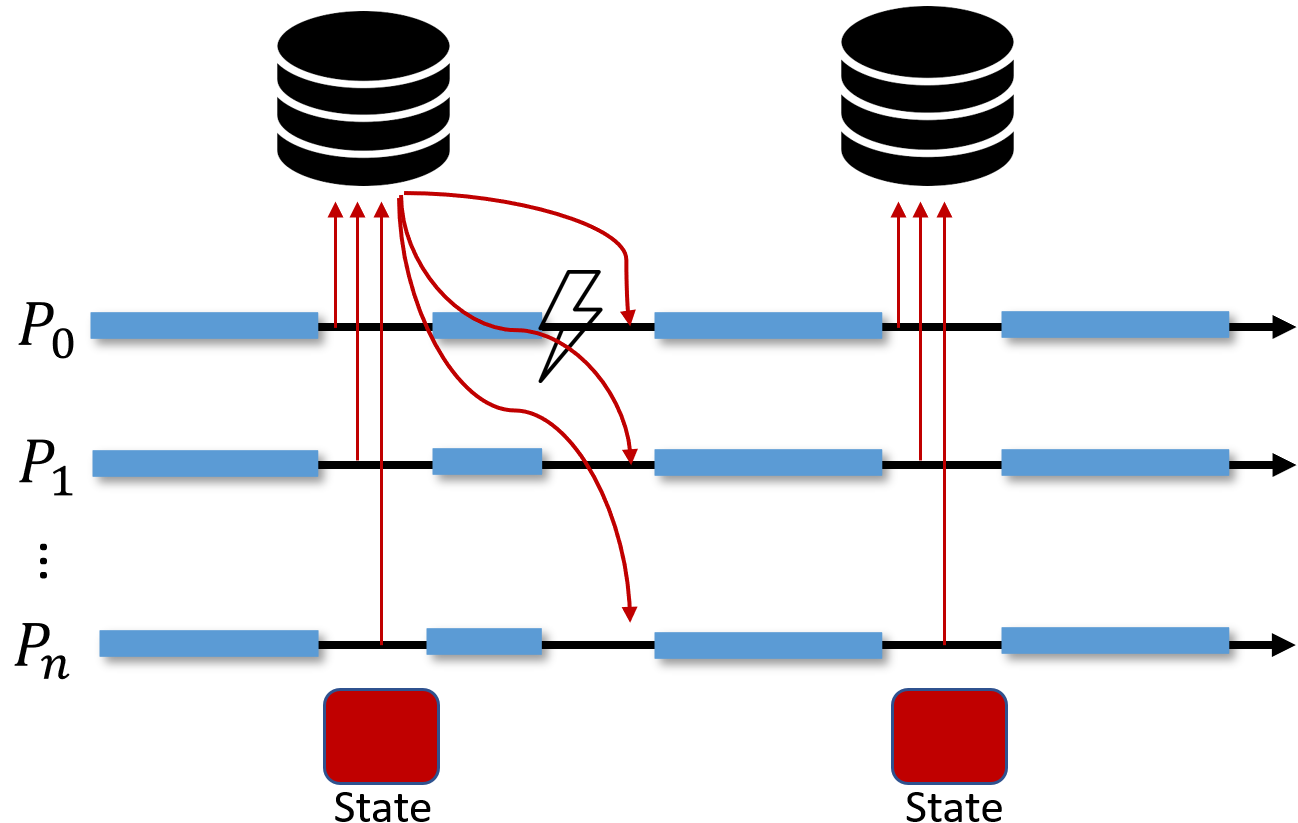}
  \caption{CR to Storage}
  \label{fig:cr-storage-model}
\end{subfigure}%
\begin{subfigure}{0.25\textwidth}
  \centering
  \includegraphics[width=4.2cm,height=3cm]{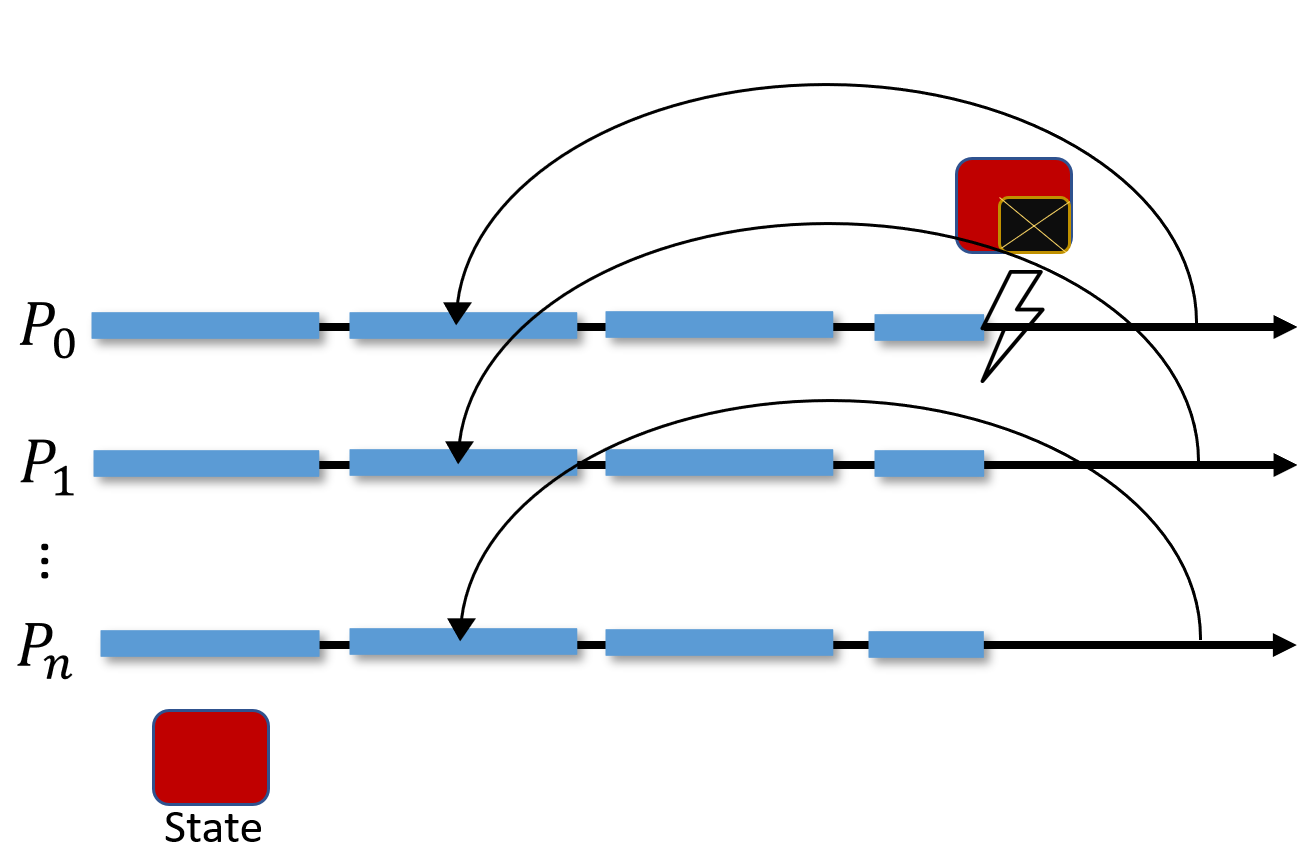}
  \caption{Intrinsic Recovery Mechanisms}
  \label{fig:intrinsic-recovery-model}
\end{subfigure}%
\begin{subfigure}{.25\textwidth}
  \centering
  \includegraphics[width=4.2cm, height=3cm]{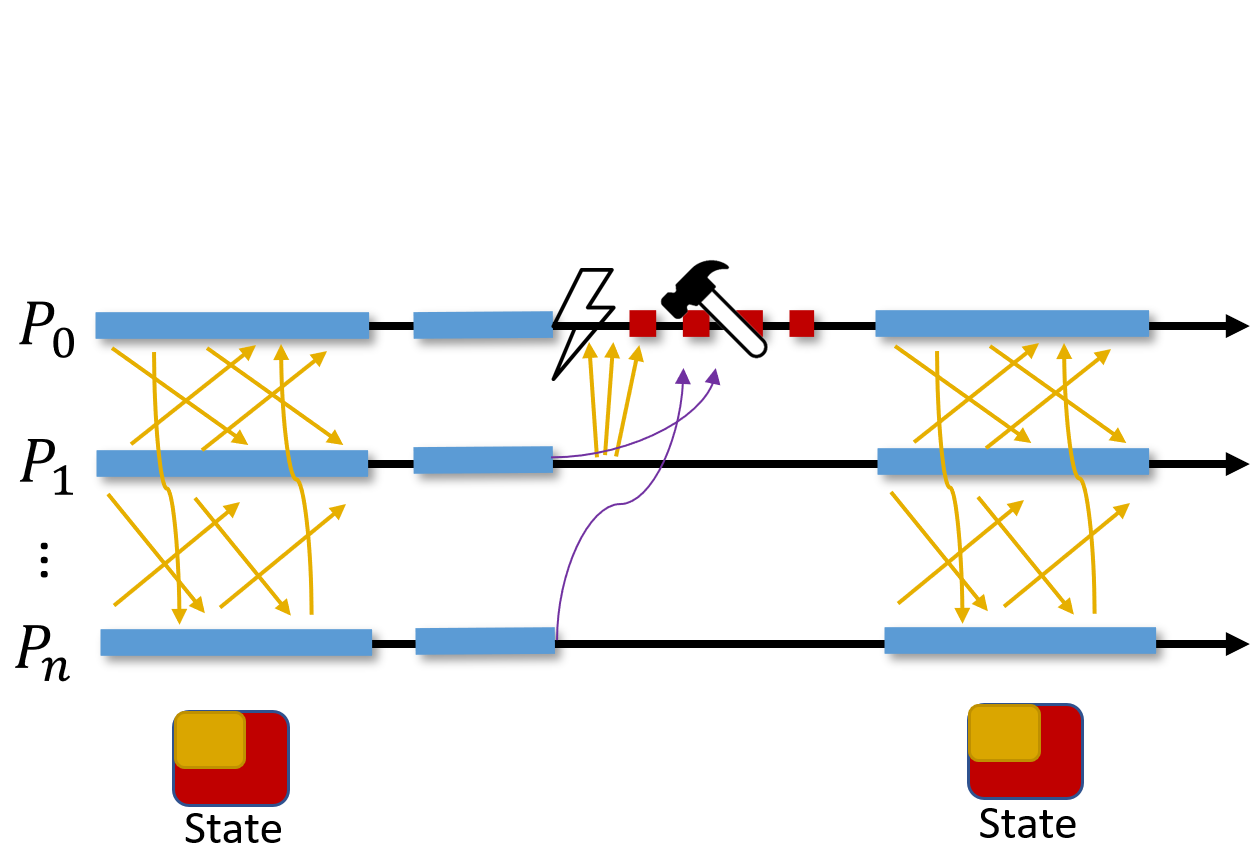}
  \caption{\textit{In-RAM} ESR}
  \label{fig:in-ram-esr-model}
\end{subfigure}%
\begin{subfigure}{.25\textwidth}
  \centering
 \includegraphics[width=4.2cm,height=3cm]{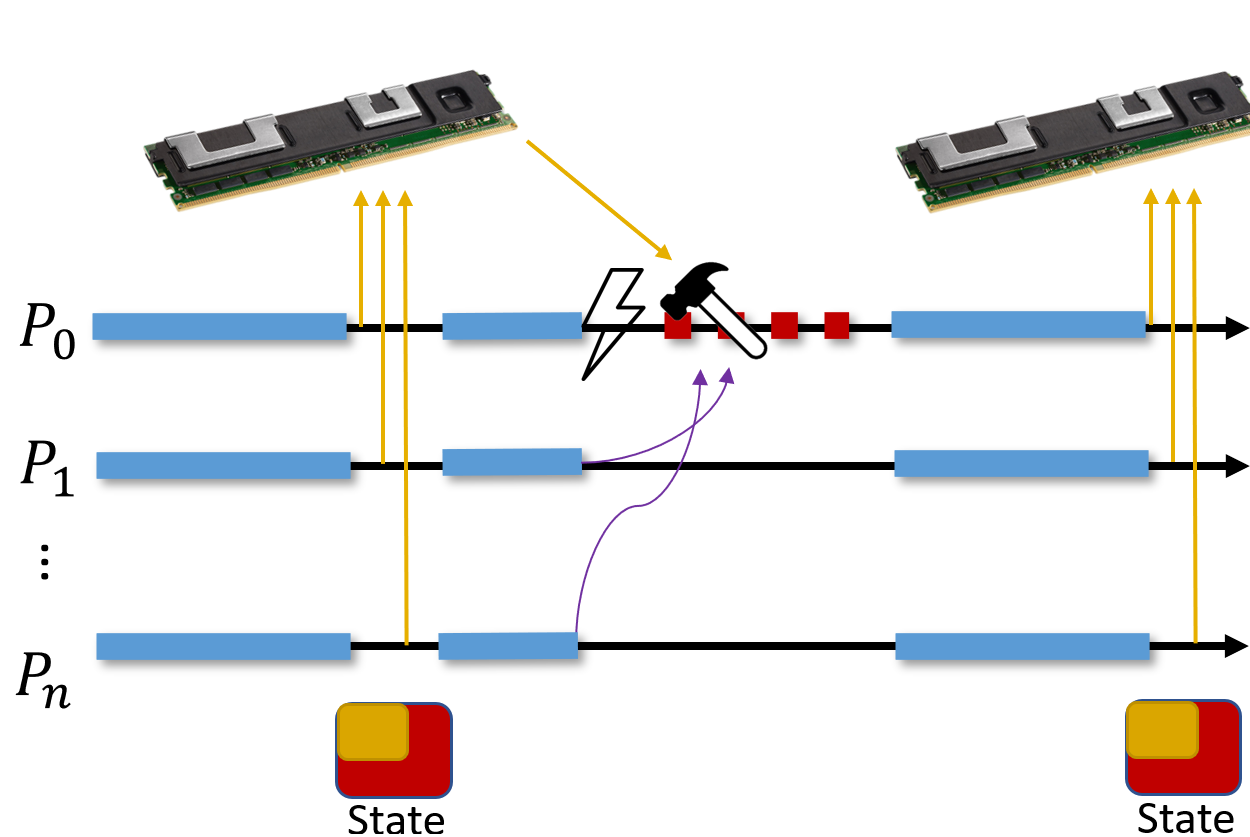}
  \caption{\textit{In-NVRAM} ESR}
  \label{fig:in-nvram-esr-model}
\end{subfigure}%
    \caption{Recovery models for iterative solvers: (a) 
    Checkpointing the state of the computation frequently to storage device; when a failure occurs, reading the complete state back and continuing. (b) 
    fault corrupts parts of the state, accuracy is therefore lost and recomputation is required. 
    (c) 
    Saving copies of recovery data in other processes' RAM; when a failure occurs, collecting 
    recovery data from other survived processes via messages, 
    reconstructing the full state and continuing 
    (d) 
    Exact state reconstruction with saving recovery data to NVRAM directly; when a failure occurs, 
    reading recovery data from NVRAM, reconstructing the full state and continuing.}
    \label{fig:recovery_models}
    
\end{figure*}
\begin{table*}[t]
\centering
{\scalebox{.96}{
\hspace{-0.4cm}
\begin{tabular}{ |c|c|c|c|c| } 
 \hline
   & \multicolumn{1}{c|}{\textbf{CR to Storage}} & \multicolumn{1}{c|}{\textbf{Intrinsic Recovery Mechanisms}} & \multicolumn{1}{c|}{\textbf{\textit{In-RAM} ESR}} & \multicolumn{1}{c|}{\textbf{\textit{In-NVRAM} ESR}} \\ 
  \hline
   & direct recovery, no extra calculations & minimal memory footprint and & exact state reconstruction of failed & exact state reconstruction with marginal \\ 
  \multirow{3}{*}{\rotatebox[origin=c]{90}{\parbox[c]{0cm}{\centering \textbf{Pros}}}} & are needed to reconstruct the & manipulation of recovery data & processes &  memory footprint of recovery \\ & computation &  &  & data \\
  \hline
   & frequent checkpoints to standard & reached state is lost, recovery & huge memory and networking & NVRAM performances are still not \\
   \multirow{3}{*}{\rotatebox[origin=c]{90}{\parbox[c]{0cm}{\centering \textbf{Cons}}}} & SSD or HDD are expensive  &  requires extra calculations to make  & overheads &  comparable to RAM performances \\ 
   & (in memory and time)  & up for the lost accuracy & & \\ 
 \hline 
\end{tabular}
}} 
 
\caption{Prominent Pros and Cons of the various recovery models described in Fig.~\ref{fig:recovery_models}.}
\label{table:pros_cons_recovery_models}
\end{table*}

Yet another recovery scheme is to design  recovery algorithms that tolerate failures according to the specific characteristics of an application \cite{arnold2008reliable,chen2006algorithm,chen2008algorithm,chen2006scalable,chen2008extending,davies2010fault,hakkarinen2010algorithmic}. A key example of such a scheme is \emph{exact state reconstruction} (ESR). ESR is applicable to many distributed linear iterative solvers. Introduced by Chen et al.~\cite{chen2006algorithm,ChenHPDC2011}, and refined by Pachajoa et al.~\cite{pachajoa2018extending,pachajoa2020algorithm,pachajoa2019make,pachajoa2020generic}, ESR is a  recovery mechanism that achieves exact reconstruction of the computation state under failures while incurring significantly lower overhead than CR \cite{pachajoa2018extending}.
ESR provides fault tolerance by keeping redundancies of chosen process'
state variables in the memory of other processes in run time; we therefore 
refer to it as \textit{in-RAM} ESR.\footnote{
We use the terminology \textit{in-X}, where \textit{X} is the target of the redundancy needed for the ESR operation. 
Specifically, \textit{in-RAM} (as in the original ESR), \textit{in-SSD} and \textit{in-NVRAM.} 
We note that the target device does not directly imply the access fashion. 
Specifically, while RAM will be byte-addressable in the context of this paper, 
NVRAM can be either byte- (with \textit{DAX} and RDMA operations, or byte-oriented PMFS) 
or block-addressable (with simple PMFS). 
Of course, SSDs and HDDs will be referred as block-oriented only.}. 
Chen~\cite{ChenHPDC2011} introduced a way to use the \emph{sparse matrix–vector multiplication} 
(\emph{SpMV}) to redundantly store the input vector, 
exploiting the already existing transmission of data between 
processes. In this manner, redundant copies of the input vector can be 
produced with relatively low memory and runtime overheads.
Pachajoa et al.~\cite{pachajoa2018extending, pachajoa2019make, pachajoa2020algorithm} 
extended \textit{in-RAM} ESR to support multiple node failures. 
However, to maximize fault tolerance, processes replicate their state vectors on each other compute node which comes with a significant memory expense. Additionally, in this case redundancies are sent all-to-all after each iteration (or after a certain period), leading to a surge in network traffic. To alleviate these problems, ESR can replicate the state at only a fraction $\alpha$ of the cluster, for example, at half of the nodes ($\alpha=0.5$). It is also possible to use different strategies for selecting the nodes to keep the redundant copies for minimizing communication overheads during SpMV~\cite{pachajoa2019make}. Nevertheless, the original problem remains as the scale of computers grows (see Section~\ref{sec:ESR memory usage}), leaving open the challenge of achieving full resilience while reducing the costs of ESR and improving its scalability. 
\subsection{Our contribution: \textit{In-NVRAM} ESR}
In this work we investigate how to significantly improve the performance of \textit{in-RAM} ESR.
Specifically, we show how its extended memory footprint and network traffic can be dramatically reduced.
Our work rests on three pillars: (1) recently enabled capabilities of \textit{direct access (DAX)} to NVRAM, (2) the access to such memory using MPI One-Sided Communication (OSC) over RDMA, and (3) the observation that these two capabilities allow maintaining all of the advantages of the original \textit{in-RAM} ESR while persisting only a single copy of recovery data during each persistence cycle, instead of maintaining multiple redundancies.
This yields the enhanced \emph{in-NVRAM} ESR, which instead of relying on and populating the RAM with many redundancies for fault tolerance, sends just a single copy DAX-wise through RDMA directly to the persistent NVRAM. 
Accessing byte-addressable NVRAM directly, without incuring the latency of moving data to and from the I/O bus, with comparable performances to RAM, and while incurring only a small overhead, yields an enhanced ESR mechanism, without compromising data and recovery consistency. 



We implement \textit{in-NVRAM} ESR using the extension of 
MPI One-Sided Communication (OSC) over RDMA \cite{dorozynski2016checkpointing,mpi_win_ext} under the setting of NVRAM. We study two possible NVRAM placements architectures: 
\begin{enumerate}
\item \emph{Homogeneous NVRAM cluster}, 
in which each compute node is equipped with its own NVM module, 
enabling the persistence of ESR state variables to local NVRAM 
by using either the \emph{persistent memory development kit} (\emph{PMDK})~\cite{scargall2020introducing} libraries or a local MPI window.
\item \emph{NVRAM persistent recovery data (PRD) sub-cluster}, in which recovery data is persisted 
in dedicated PRD sub-cluster nodes via remote MPI 
one-sided communication implemented using RDMA. 
\end{enumerate}
In the PRD sub-cluster architecture, we assume RAID between nodes to provide fault tolerance to errors in the sub-cluster. Otherwise, each node of the sub-cluster behaves as a single point of failure. We stress that while \textit{in-RAM} ESR's data transportation increases quadratically with the cluster size (as explained in Section~\ref{sec:ESR memory usage}), increase in writes for RAID is linear and depends on RAID level. 

We note that, to the best of our knowledge, planned exascale supercomputers are expected to use the remote PRD NVRAM sub-cluster architecture~\cite{auroraoverview,auroraengineering,io500list}. 
This choice of architecture is mainly because integrating 
Optane DCPMM modules in each compute node 
would reduce the number of available DDR DIMM memory slots, 
thereby reducing the size of available DRAM, 
which is one of the crucial resources in HPC nodes~\cite{li2018performance}. 
Hence, future supercomputers integrating Optane DCPMM modules are
designed with remote NVM storage nodes, 
e.g., the DAOS storage server~\cite{hennecke2020daos} in Aurora~\cite{Argonne1,auroraoverview,auroraengineering}.
Nevertheless, this work evaluates also the homogeneous NVRAM cluster architecture, 
as future systems might include NVRAM in each compute node,
for fast storage or relatively cheap and fast byte-addressable memory expansion 
(see \cite{fridman2021assessing}, \cite{patil2021nvm}). Specifically, network and remote I/O performance characteristics play a crucial role in the optimization of HPC applications running in HPC cloud systems (HPCaaS). Thus, future HPC cloud systems are integrating storage devices such as NVMe-based SSD locally~\cite{kashyap2022nvme,guz2018performance,gupta2011evaluation}, and even foundations for NVRAM integration are already established \cite{https://doi.org/10.48550/arxiv.2208.02240}. 

Fig.~\ref{fig:recovery_models} describes the different types of recovery models for 
iterative solvers, while Table~\ref{table:pros_cons_recovery_models} 
summarizes their pros and cons. 


In summary, our work incorporates NVRAM recovery with exact state reconstruction 
techniques for distributed linear iterative solvers. 
We propose, implement and evaluate \textit{in-NVRAM} ESR, a novel NVRAM-based mechanism 
for scalable and resource-efficient exact state reconstruction techniques.
We implemented \textit{in-NVRAM} ESR in the PRD sub-cluster architecture 
by using MPI one-sided communication over InfiniBand's remote direct memory access (RDMA) 
towards NVRAM and have optimized its usage by ESR's persistence iterations (see Section~\ref{sec:implementation OSC-RDMA}).

To the best of our knowledge, our work is the first to report on a scientific application implementation that accesses remote NVRAM in this manner. 

We conducted a comprehensive performance evaluation, comparing \textit{in-NVRAM} ESR 
to other implementations that store recovery data, 
either in DRAM (\textit{in-RAM ESR}) or in SSD storage device (\textit{in-SSD ESR}). 
Our evaluation shows that \textit{in-NVRAM} ESR is significantly superior to \textit{in-RAM} 
ESR in terms of the size of the memory footprint required for storing the recovery data. 
\textit{In-NVRAM} ESR is also superior to \textit{in-RAM} ESR and \textit{in-SSD} 
ESR in terms of the time overhead incurred by writing the recovery data. 
Based on these results, we estimate that the small memory and time overheads 
of \textit{in-NVRAM} ESR will allow it, unlike ESR, to be deployed in 
future exascale supercomputers for providing high resiliency to the important class 
of distributed iterative solver for linear systems algorithms for which ESR is suitable.  

\paragraph*{Organization}
The rest of the paper is organized as follows.
 Section~\ref{esr} describes the \textit{in-RAM} ESR technique and its challenges for the scalability of linear iterative solvers to exascale systems. In section~\ref{in-NVRAM ESR} an NVRAM-based solution is described and the \textit{in-NVRAM} ESR model is suggested, applicable to the homogeneous NVRAM cluster architecture and the PRD sub-cluster architecture. In section \ref{sec:ESR memory usage} we compare the memory utilization of \textit{in-RAM} ESR and \textit{in-NVRAM} ESR and demonstrate significant savings in memory resources when persisting data to NVRAM with \textit{in-NVRAM} ESR. In section \ref{sec:implementation} we describe the implementation details of \textit{in-NVRAM} ESR, focusing on the technologies and software that enable \textit{in-NVRAM} ESR to access NVRAM locally and remotely. Section \ref{key} presents the evaluation results of \textit{in-NVRAM} ESR in a  key example of the widely used PCG solver for sparse systems. 

\section{\textit{In-RAM} ESR and Its Challenges}\label{esr}

\emph{Exact state reconstruction} (\textsc{ESR}) is a technique for recovering the state of a linear algebra iterative solver after a failure, avoiding the checkpointing of the entire state of the computation. ESR takes advantage of concurrent data distributed between nodes to reconstruct the state~\cite{pachajoa2019make,levonyak2020scalable}. Specifically, it exploits \emph{sparse matrix-vector multiplication} (\emph{SpMV}) operations to produce redundant copies to vectors with low memory and runtime overheads. Therefore, ESR is applicable to iterative solvers that involve \emph{SpMV} and perform a finite-term recurrence (hence the state can be reconstructed from a bounded amount of previously calculated data).

ESR first identifies the state of the solver. 
Upon recovery, the redundancy of vectors participating in the SpMV operations
is used to reconstruct the other state vectors, 
by solving local equations on a replacement node. When the full state of the failed process is reconstructed, the computation can proceed on the replacement node. In ESR, whenever the SpMV operation is applied, \textit{i.e.}, $Av^{(j)}$ is computed for some state variable $v$ at iteration $j$, the transition of $v$-values is augmented to create redundancy for all its entries. This augmented SpMV operation is denoted by ASpMV~\cite{pachajoa2020algorithm}. Redundancies are created on other processes' RAM, hence this model is referred to as \textit{in-RAM} ESR (see Fig.~\ref{fig:volatile_cluster}). 

A generic method for this state identification for iterative solvers is described in by Pachajoa et al.~\cite{pachajoa2020generic}. 
This meta-algorithm uses the dependency graph of an iterative algorithm to decide which variables have to be saved and which can be reconstructed by using other variables. 
Two main examples demonstrated are the PCG ~\cite{nazareth2009conjugate} 
and the BiCGStab~\cite{sleijpen1994bicgstab} solvers. 
We stress that these two solvers are examples of a larger class of iterative solvers that can be adjusted to ESR. 
For example, in Jacobi, SOR and Gauss-Seidel the solution approximation variable will keep redundancies, 
and in MINRES and GMRES it will be the Arnoldi vectors~\cite{saad1986gmres}. 

For a given distributed linear iterative solver produced with an ESR algorithm, we denote the variables of the full state by the set V, and the subset of ESR variables that participate in SpMV operations and create redundancies by $V_{SpMV}$ ($V_{SpMV} \subseteq V$). 
We denote by $n$ the size of each global vector of the solver. We refer to the set of all indices as $I=\{1 \dots n\}$. The indices corresponding to a certain process $p$ are denoted by $I_p$. 
Specifically, $f$ denotes a failed process, and its indices are denoted $I_f$. The value of variable $v$ at the $j^{th}$ iteration is denoted by $v^{j}$. \\

In the reconstruction phase of failed process $f$, 
the redundancies of $(V_{SpMV})_{I_f}$ are collected to 
a replacement node, together with the values 
of $V_{I\setminus I_f}$ and $V_{I\setminus I_f}$ that are needed for reconstruction from the surviving nodes. For some multi-term recurrence solvers, reconstruction requires $k$ successive values of $(V_{SpMV})_{I_f}$, and therefore redundancies for such variables are kept for $k$ iterations. For example, in the ESR algorithm for the Preconditioned Conjugate Gradient (PCG) solver, presented in \cite{ChenHPDC2011}, $k=2$ as it stores the last two search directions via ASpMV (the PCG solver is a two-term recurrence). Solving some local linear equations, $(V\backslash V_{SpMV})_{I_f}$ is reconstructed, ending up with the full state of the process, which enables to continue the computation.


To tolerate multiple node failures, the redundancy should be saved 
in multiple copies.
Thus, even when several nodes crash together, values can still be recovered 
from the RAM of processes on the surviving nodes. 
If $c$ nodes may fail simultaneously, $c$ redundant copies should be made. Copies should be saved on different nodes, since if a node crashes, all of its processes fail together. In this case, $I_f$ represents the indices of all the failed processes together, 
and the reconstruction algorithm is executed on several nodes, 
solving the local equation systems together distributively. The reconstruction effort depends on the number of failed processes. As the number of processes that fail simultaneously increases, the size of the equations that must be solved for reconstructing the lost data grows.

There is a delicate balance between the runtime overhead 
required to save the recovery data of the application securely, and the time it takes to recover after a failure. ESRP~\cite{pachajoa2020algorithm} is a modification of ESR, where redundant copies are created periodically to alleviate the networking overhead for each iteration. 
ESRP demonstrates a trade-off, where 
increasing the period of ESR decreases the runtime overhead, 
but increases the cost of discarding the iterations performed 
since the last copies were made when recovery is required. 

ESR aims to minimize the amount of data being persisted, 
by a careful analysis of the application that identifies 
a minimum-sized set of variables whose state should be made persistent. 
These variables must be chosen such that all other significant variables can be reconstructed from their values. 
A generic method for this state identification for iterative solvers is
described in~\cite{pachajoa2020generic}. This generic meta-algorithm produces ESR algorithms for various solvers. Two main examples demonstrated by Pachajoa are the PCG~\cite{nazareth2009conjugate} and the BiCGStab~\cite{sleijpen1994bicgstab} solvers. We stress that these two solvers are examples of a larger class of iterative solvers that can be adjusted to ESR. For example, in Jacobi, SOR and Gauss-Seidel the solution approximation variable will keep redundancies, and in MINRES and GMRES it will be the Arnoldi vectors~\cite{saad1986gmres}. 

\textit{In-RAM} ESR has three main limitations. First, a large number of copies should be made to redundant data to ensure the recovery when a large number of nodes fail simultaneously. This creates a dramatic increase in memory overhead, and therefore effectively reduces the size of the problems that can be solved. Second, \textit{In-RAM} ESR suffers from networking and time overheads, as the redundancies are sent between nodes simultaneously --- leading to a surge in network traffic. Finally, \textit{In-RAM} does not rely on NVRAM, which is a main disadvantage for future supercomputers that are planned to integrate NVRAM. NVRAM-based systems offer an attractive alternative solution to 
these challenges, as we describe next. 

\section{\textit{In-NVRAM} ESR: Overview}\label{in-NVRAM ESR}

To exploit NVRAM for ESR recovery, we use it for persisting the redundant data required for it instead of keeping it in the memory of other nodes. We name this mechanism \emph{in-NVRAM} ESR. 
In NVM-based systems, processes have access to 
persistent memory space, in addition to a unique volatile memory space available for every process. Nodes crash arbitrarily and independently of each other. Upon a crash of a certain node, the content of its volatile memory is lost, and the data in its persistent memory (if it includes NVM) becomes inaccessible until the node recovers.

\begin{figure*}
\vspace{-0.7cm}
\begin{subfigure}{0.27\textwidth}
  \centering
  \includegraphics[width=4.5cm,height=4cm]{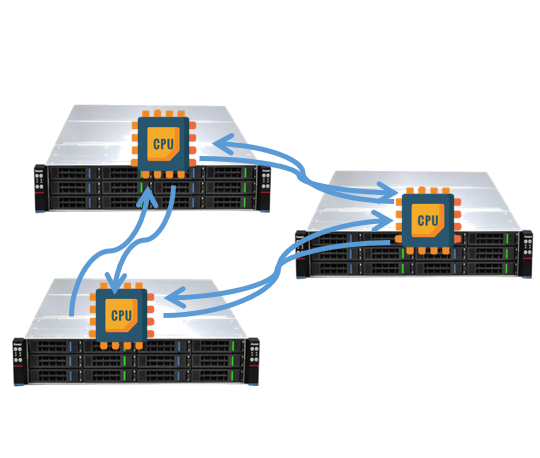}
  \label{fig:volatile_cluster}
\end{subfigure}%
\begin{subfigure}{.37\textwidth}
  \centering
  \includegraphics[width=5cm, height=4cm]{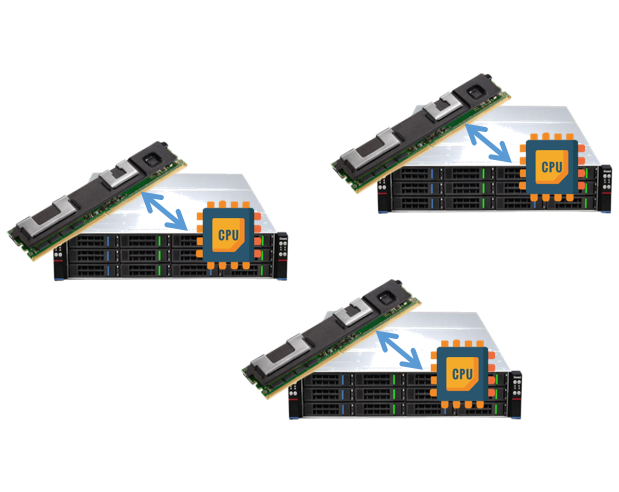}
  \label{fig:homogeneous_cluster}
\end{subfigure}%
\begin{subfigure}{.33\textwidth}
  \centering
 \includegraphics[width=3.8cm,height=4cm]{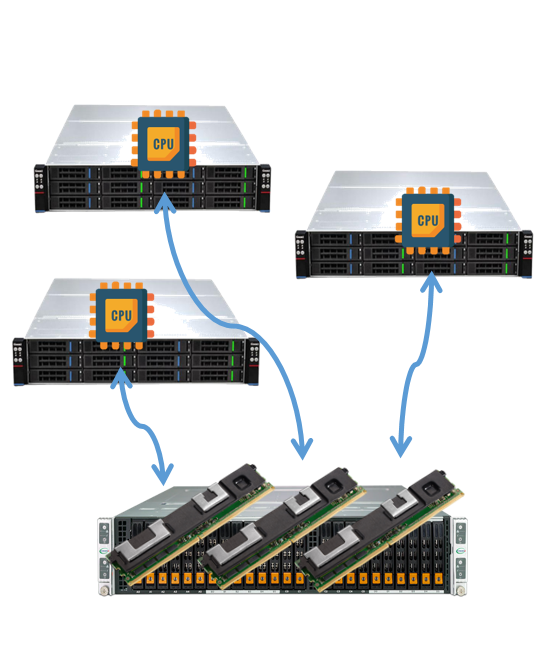}
  \label{fig:storage_cluster}
\end{subfigure}%
\\
\vspace{-0.2cm}
\begin{subfigure}{0.33\textwidth}
  \centering
  \includegraphics[width=4.5cm,height=3cm]{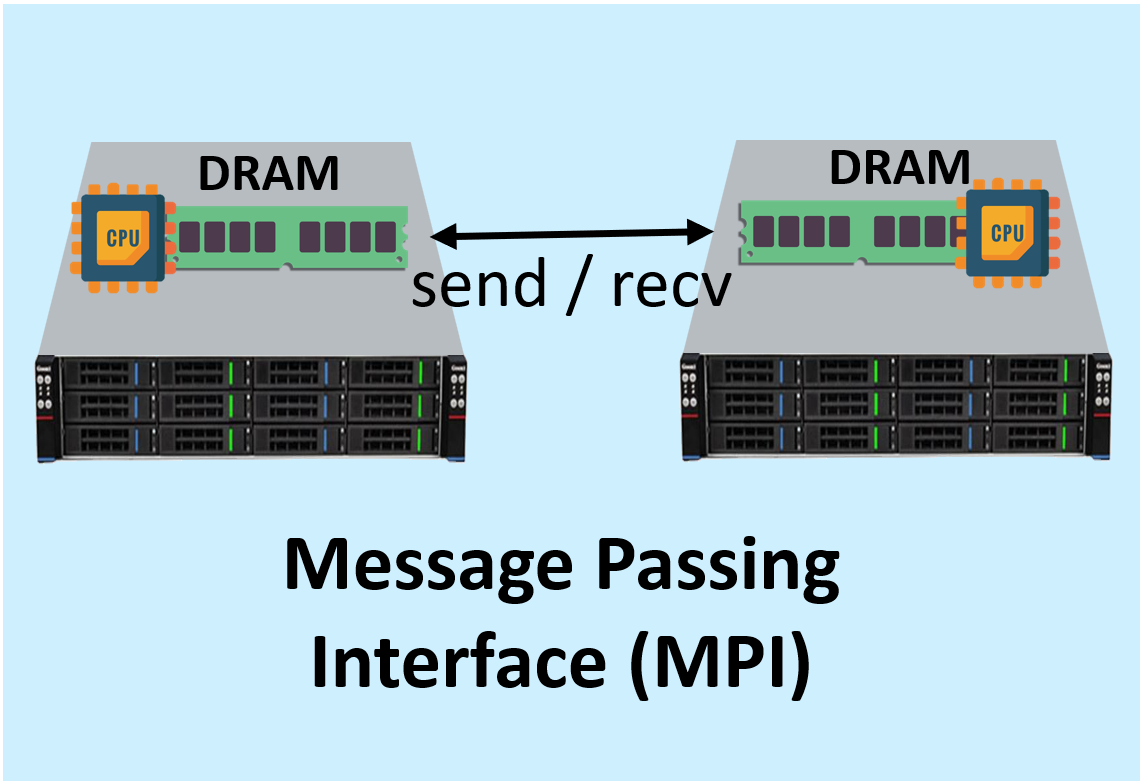}
  \caption{Volatile memory architecture.}
  \label{fig:volatile_cluster}
\end{subfigure}%
\begin{subfigure}{.33\textwidth}
  \centering
  \includegraphics[width=4.5cm, height=3cm]{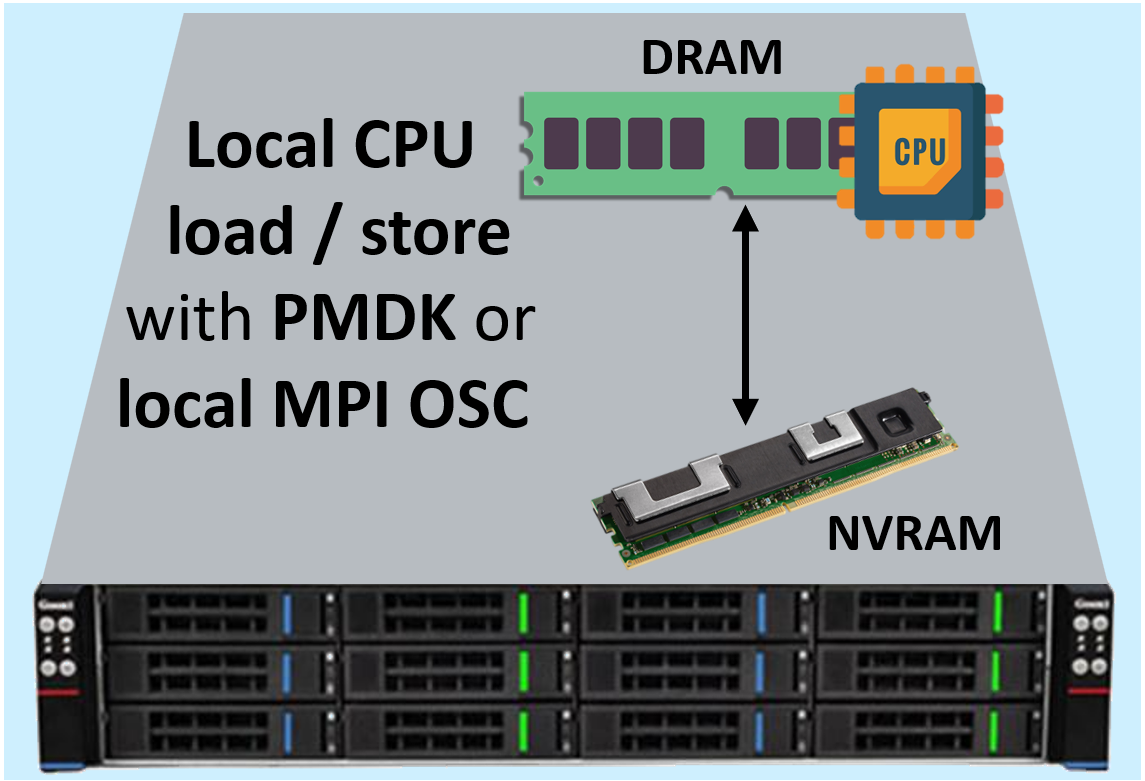}
  \caption{Homogeneous NVRAM cluster.}
  \label{fig:homogeneous_cluster}
\end{subfigure}%
\begin{subfigure}{.33\textwidth}
  \centering
 \includegraphics[width=4.5cm,height=3cm]{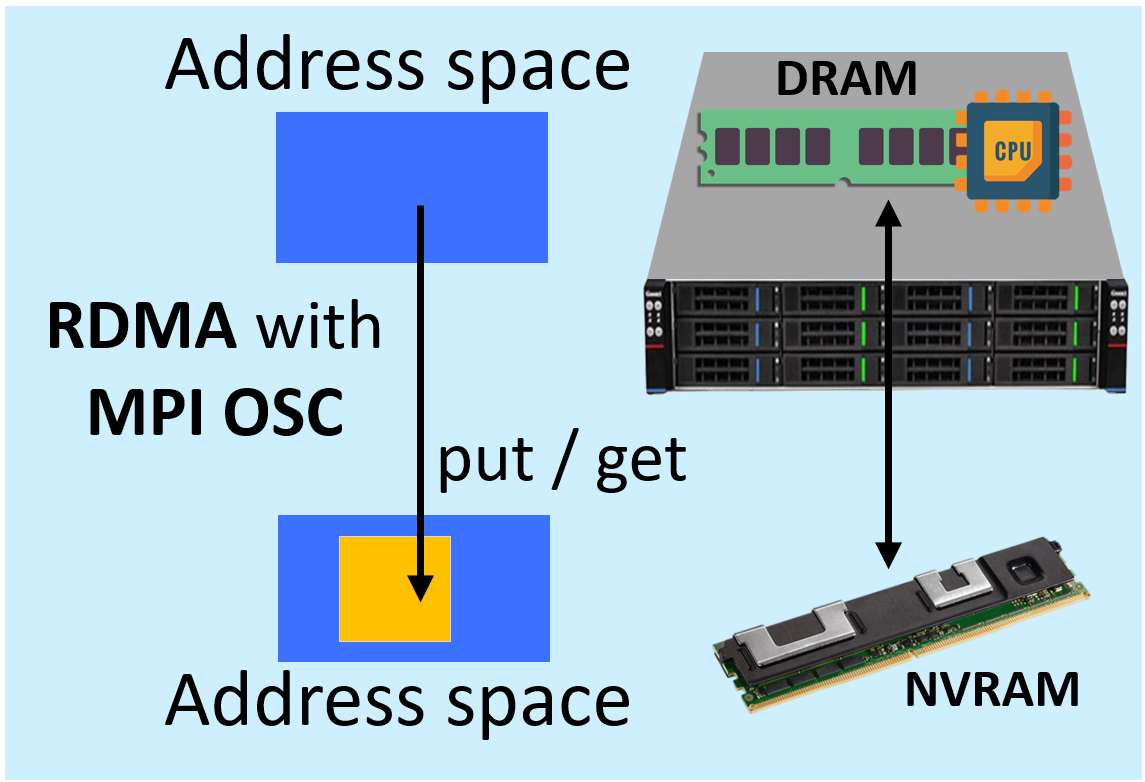}
  \caption{Persistent recovery data (PRD) sub-cluster.}
  \label{fig:storage_cluster}
\end{subfigure}%
    \caption{Cluster architectures with and without the usage of NVRAM.}
    \label{fig:nodes}
\end{figure*}
There are two key possible ways to integrate NVRAM devices into a 
cluster architecture. 
In the first, more traditional architecture, 
called \textit{homogeneous NVRAM cluster},
each node has an NVRAM device attached to it (see Fig.~\ref{fig:homogeneous_cluster}).
Each node's state is saved into its local NVRAM whose coherency 
should be ensured by the application. 
If a node fails, the persisted data can be recovered from the NVM once it recovers. Therefore the reconstruction algorithm for failed process $f$ is executed on the same node that $f$ executed on before the failure, and only after it recovers.

In the second architecture, there is a sub-cluster of one or more 
\emph{persistent recovery data} (\emph{PRD}) nodes, 
each containing NVRAM modules, 
which stores recovery data for the rest of the cluster's \emph{compute nodes} 
(see Fig.~\ref{fig:storage_cluster}). In this architecture, recovery data is saved via RDMA operations with the remote memory MPI's One-sided API (to reduce overheads). (The details of how this is done are explained in 
Section~\ref{sec:implementation OSC-RDMA}.) When a process $f$ fails, its recovery data is accessible on the NVRAM sub-cluster and can be retrieved by a replacement node. For each of the failed processes, the reconstruction can be executed on every compute node that has access to the NVRAM sub-cluster. A failure of a PRD node renders its memory inaccessible, which may make it a single point of failure. This type of failure can be addressed by adding redundancy within the PRD sub-cluster itself (for example using RAID), but this is outside the scope of this paper.

The \textit{in-NVRAM} ESR persistence stage, as well as the \textit{in-NVRAM} ESR reconstruction phase for both these architectures, appear in 
Algorithm~\ref{alg:in-nvram-esr} and Algorithm~\ref{alg:in-nvram-reconstruction}, respectively. We do not deal here with the trade-off between the ESR period and the number of ``wasted'' iterations upon a failure demonstrated
in ESRP~\cite{pachajoa2020algorithm} and simply assume that 
this period is chosen optimally according to the problem and 
the cluster characteristics. We focus instead on a single persistence iteration during the calculation.
\begin{algorithm}[b]
\caption{\textit{In-NVRAM} ESR for a persistence iteration $j$ of process $p$}
\label{alg:in-nvram-esr}
\begin{algorithmic}
\State Compute $j^{th}$ iteration of the solver
\Statex \vspace{-0.1cm} \hspace{0.4cm}\vdots
\Statex \textbf{if} \textsc{homogeneous Cluster} \textbf{then}
\Statex \ \ \ \  persist $(V_{SpMV})^j_{I_p}$ to \textbf{local} NVM
\Statex \textbf{if} \textsc{PRD sub-cluster} \textbf{then}
\Statex \ \ \ \  persist $(V_{SpMV})^j_{I_p}$ to \textbf{remote} NVM
\end{algorithmic}
\end{algorithm}

\begin{algorithm}[h]
\caption{\textit{In-NVRAM} ESR Reconstruction phase of iteration j of failed process $f$}
\label{alg:in-nvram-reconstruction}
\begin{algorithmic}
\Statex \hspace{-0.6cm} \textbf{if} \textsc{homogeneous Cluster} \textbf{then} 
\Statex \hspace{-0.6cm} \ \ \ \ Wait for failed nodes to recover and have access to \textbf{local} NVM
\Statex \hspace{-0.6cm} \textbf{if} \textsc{PRD sub-cluster} \textbf{then}
\Statex \hspace{-0.6cm} \ \ \ \ Run reconstruction from any spare nodes that have access to \textbf{remote} NVRAM Sub-Cluster 
\vspace{-0.15cm}
\Statex \hspace{-0.6cm} \hrulefill
\Statex Retrieve the static data of the solver
\Statex Gather required variables from $V_{I\setminus I_f}^{(j)}$
\Statex \textbf{if} \textsc{homogeneous Cluster} \textbf{then}
\State \ \ \ \ Read $\{(V_{SpMV})^i_{I_f}\}^{j}_{i=j-k+1}$ from \textbf{local} NVM
\Statex \textbf{if} \textsc{PRD sub-cluster} \textbf{then}
\State \ \ \ \ Read $\{(V_{SpMV})^i_{I_f}\}^{j}_{i=j-k+1}$ from \textbf{remote} NVM
\Statex \vspace{-0.2cm} \hspace{0.4cm}\vdots
\State Solve local linear systems to reconstruct $(V\backslash V_{SpMV})^j_{I_f}$ 
\end{algorithmic}
\end{algorithm}
\subsection{Comparing \textit{in-RAM} ESR and \textit{in-NVRAM} ESR}
\label{sec:ESR memory usage}
Let $N$ denote the number of compute nodes;
The number of the actual compute processes used by the solver is  
$proc\leq t \cdot N$.
$M_{R}(n, proc, \phi)$ denotes the amount of memory overhead required 
by a recovery method $R$ to support the recoverability of up to $\phi$ simultaneous node failures. We assume that the vulnerability of the system is proportional to the number of nodes, hence $\phi=\alpha \cdot N$ for some constant $\alpha<1$. In the following, we estimate
$M_{R}(n, proc, \phi)$, where R=\{\textit{in-RAM} ESR, \emph{in-NVRAM} ESR\}.
In the \textit{in-RAM} ESR mode, redundancies for the variables in $V_{SpMV}$ are saved for successive $k$ iterations in $\phi$ nodes, therefore 
\begin{align*}
M_{\textit{in-RAM}\text{ ESR}}(n, proc, \phi)=\Sigma_pk\cdot|(V_{SpMV})_{I_p}|\cdot \phi \\
        =k\cdot|V_{SpMV}|\cdot \phi=O(|V_{SpMV}|\cdot \phi)
\end{align*}
elements in memory. In the \textit{in-NVRAM} ESR mode, a single copy (up to certain RAID level) is saved in NVRAM for the variables in $V_{SpMV}$, therefore 
\begin{align*}
M_{\textit{in-NVRAM}\text{ ESR}}(n, proc, \phi)= O(\Sigma_pk\cdot|(V_{SpMV})_{I_p}|) \\
    =O(|V_{SpMV}|)
\end{align*}
elements in persistent area (where the $O()$ notation hides the constant of RAID level). 

For sparse linear problems of matrices of size $n \times n$, the representation of the problem is linear in $n$. In addition, the size of the problem is estimated to be proportional to $N$ (as more nodes add more RAM). Moreover, $|V_{SpMV}|$ is proportional to $n$ as $V_{SpMV}$ consists of global vectors of size $n$. Therefore we conclude that $|V_{SpMV}|$=$O(N)$. Remember that typically $\phi$ is chosen in proportion to $N$ ($\phi=\alpha\cdot N$). Finally, we write $M_{\textit{in-RAM}\text{ ESR}}(n, proc, \phi)=O(N\cdot N)=O(N^2)$, while  $M_{\textit{in-NVRAM}\text{ ESR}}(n, proc, \phi)=O(N)$. 
These estimations demonstrate that \textit{in-NVRAM} ESR is much more scalable than \textit{in-RAM} ESR as the former incurs fault tolerance overheads in memory that increase linearly with the cluster size, while the latter incurs memory overheads that increase quadratically. Moreover, while \textit{in-RAM} ESR uses the RAM to save the huge amount of recovery data, \textit{in-NVRAM} ESR uses persistent memory for its recovery data, leaving more space in RAM for larger computations.  

\subsection{Comparing \textit{In-NVRAM} ESR and \textit{In-SSD} ESR}
Checkpointing scientific applications, either transparently (for example with DMTCP \cite{ansel2009dmtcp}) or explicitly (for example with SCR \cite{moody2010design}), persists all the data allocated by the application to a block device, so restarting is enabled by reading the last available checkpoint into the application's buffers. Usually, and especially for distributed linear iterative solvers, not all the data is necessary for exact reconstruction, as the exact state can be reconstructed from only a partial checkpoint. ESR alleviates the full state's checkpoint by persisting only a subset of the state and computationally reconstructing the rest of the state upon recovery. \\
\textit{In-SSD} ESR uses a block SSD device to persist recovery data. This includes all the I/O stages, from system call invocation, block granularity writes, and data transferring via the I/O bus. All these add additional overhead to the high latency of block SSD devices. 
In contrast, the \textit{in-NVRAM} ESR mechanism persists the recovery data to the NVRAM directly, either locally via the memory bus or remotely via RDMA. Unlike block-based systems, this eliminates the intervention of the operating system by directly accessing the byte-addressable NVRAM. Accessing the NVRAM in this manner is much faster and, in fact, persistence operation only access the application's address space, unlike when accessing SSD.

\section{The Implementation in Detail}
\label{sec:implementation}

This section describes the various ways in which \textit{in-NVRAM} ESR can 
use the system's capabilities to persist data to NVRAM. 

\subsection{MPI One-Sided Communication (OSC) over RDMA}
\label{sec:implementation OSC-RDMA}

Remote direct memory access (RDMA) support in networks became mainstream, 
particularly through the widespread adoption of InfiniBand as 
a commodity network fabric~\cite{using_advanced_mpi}. 
MPI-1 provides a powerful and complete interface for the message-passing 
approach. 
MPI-2 added (and MPI-3 greatly extended) remote memory operations
that provide a way to access the memory of remote processes directly, 
through operations that put data to, get data from, or update data at a remote process. 
Unlike message passing (using standard \textit{send} and \textit{receive} operations), 
the program running on the remote process does not need to call 
any routines to match the \textit{put} or \textit{get} operations. 
Thus, remote memory operations
can provide better performance for distributed and parallel programs. 
This functionality of remote memory access is implemented 
via \emph{memory windows}. 
The term \emph{window} is used since MPI limits what part of a process's
address space is accessible to other processes. 

Several works~\cite{dorozynski2016checkpointing,du2021fast,liu2019write}
present different persistence schemes to correctly utilize RDMA over NVRAM. 
Dorożyński et al.~\cite{dorozynski2016checkpointing} incorporate non-volatile 
RAM into wrappers over MPI One-Sided API, 
in order to provide persistence of data stored in MPI windows. 
They extend the RDMA MPI One-Sided Communication (OSC) to persist 
the data stored in windows to NVRAM. 
Data consistency is ensured by copying necessary data into a separate
location. More specifically, the data is saved into two locations alternately in order 
to have at least one proper "checkpoint" even if a failure occurs during checkpointing creation. 
An implementation of this method~\cite{mpi_win_ext} provides 
a new programming model by allowing processes to communicate freely using 
standard OSC functions and fall back to a state saved during synchronization. 
It supports synchronization calls of OSC, as explained next.

RMA communication calls of the MPI ISC API must occur in the invoking 
process only within an \emph{access epoch} for the window. 
The transferred data is available only when exiting the access epoch. 
Such an epoch starts with an RMA synchronization call on the window; 
it then proceeds with zero or more RMA communication calls 
(e.g., \texttt{MPI\_PUT} or \texttt{MPI\_GET}) on the window; 
it completes with another synchronization call on the window~\cite{mpi2015}.
RMA communications fall in two categories: 
\emph{active} and \emph{passive} target communication. 

In \emph{active target communication}, the data is moved from one process to another, 
and both processes are explicitly involved in the communication. 
In contrast to standard message passing, 
in active target communication RMA operations are controlled only 
by the invoking process, 
and the target process only participates in the synchronization. 
In active target communication, a target window can be accessed by RMA 
operations only within an \emph{exposure epoch}. 
Distinct exposure epochs at a process on the same window must be disjoint, 
but such an exposure epoch may overlap with multiple access epochs for the same window. 
MPI provides two synchronization mechanisms for active target communication: 
\begin{enumerate}
\item A general collective RMA synchronization, in which 
an access epoch at an origin process or an exposure epoch at a target 
process are started and completed by calls to \texttt{MPI\_Win\_fence}.
\item Synchronization in which only pairs of communicating processes 
synchronize using the \emph{Post-Start-Complete-Wait} (PSCW) protocol. 
In PSCW, an access epoch is started at the origin process by a call to 
\texttt{MPI\_Win\_Start} and is terminated by a call to \texttt{MPI\_Win\_Complete}. 
An exposure epoch is started at the target process by a call to 
\texttt{MPI\_Win\_Post} and is completed by a call to \texttt{MPI\_Win\_Wait}. 
The post-call has a group argument that specifies the set of origin processes 
for that epoch.
\end{enumerate}

In \emph{passive target communication}, the target process does not 
execute RMA synchronization calls, and there is no notion of an exposure epoch.
Instead, passive target synchronization is accomplished by using MPI locks at 
the origin process with \texttt{MPI\_Win\_Lock} and \texttt{MPI\_Win\_Unlock}. 
It is used to ensure that RMA operations from other processes 
do not modify the data unexpectedly.

Dorożyński et al.~\cite{mpi_win_ext} support both communication 
mechanisms in their extension of MPI OSC over NVRAM. 
For the active target synchronization, exposure epochs are closed 
with  \texttt{MPI\_Win\_Fence\_persist} (for active target communication) 
or \texttt{MPI\_Win\_Wait\_persist} (for passive target communication), 
to ensure that data reaches NVRAM before exiting the exposure epoch. 

To implement \textit{in-NVRAM} ESR in the NVRAM PRD sub-cluster architecture, 
it is necessary to ensure that data is persisted successfully to the NVRAM 
in the PRD node after each persistence iteration and before the successive 
persistence iteration attempts to access the window. 
Since the target process must know when the exposure epoch is closed,
an active target mechanism is more suitable for \textit{in-NVRAM} ESR. 

Since a persistence iteration usually requires a significant period of time, 
we can optimize by releasing the access epochs of the compute processes 
while the target process is still persisting the data in its exposure epoch.
This allows compute processes to proceed to the next compute iteration. 
For this reason, we choose the PSCW mechanism to be applied in \textit{in-NVRAM} ESR. 
Within the access epoch, a process executes a \texttt{MPI\_Win\_Put\_pmem} 
to transmit data to the remote process, 
and \texttt{MPI\_Win\_Get\_pmem} to read the data when recovery is needed. 
Whenever a compute process completes its RMA operations with \texttt{MPI\_Win\_Complete}, 
it exits the access epoch and proceeds. 
Fig.~\ref{pscw_in-NVRAM ESR} illustrates a PSCW epoch for the MPI OSC to 
an NVRAM PRD node in a persistence iteration.

A similar implementation for the homogeneous NVRAM cluster architecture, 
which we also consider, uses a separate local window for each process. 
In this architecture, each process accesses its own local window 
and persists its data locally.

\subsection{Persistent Memory Development Kit (PMDK)}
\label{sec:implementation PMDK}

\begin{figure}
\centering
\includegraphics[scale=0.175]{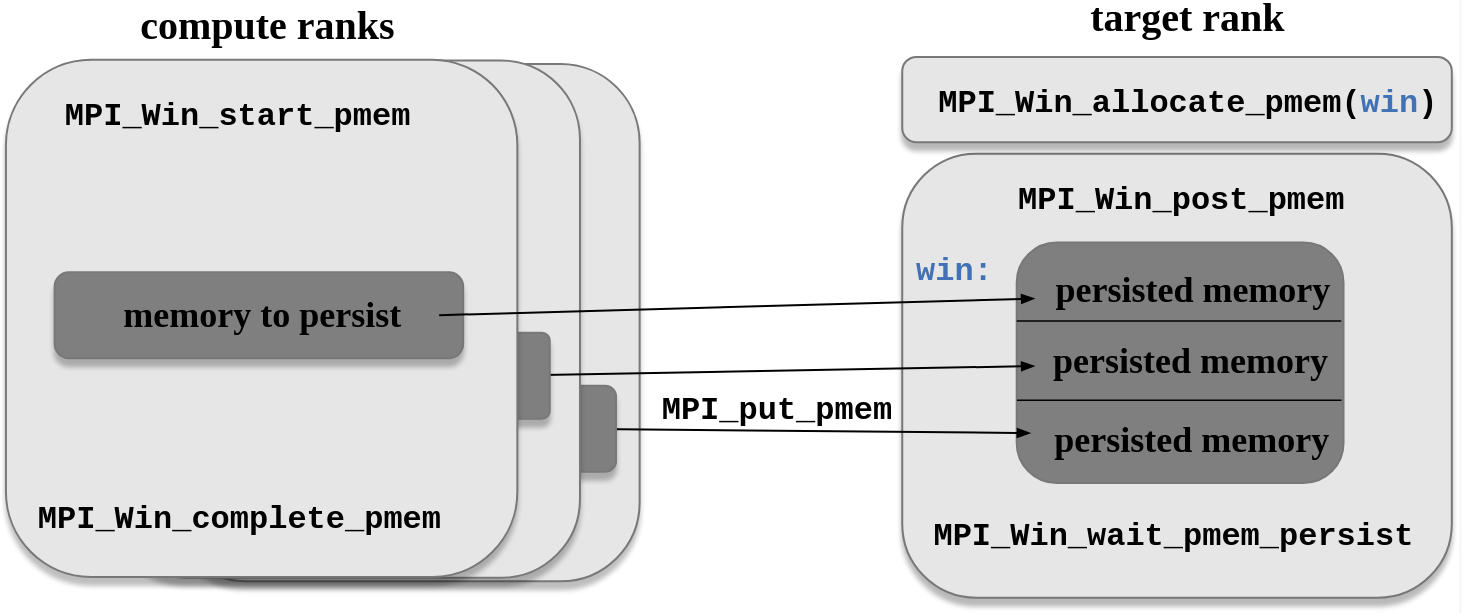}
\caption{Persisting ESR data using MPI OSC epoch over RDMA to a remote NVRAM node,
using PSCW.}
\label{pscw_in-NVRAM ESR}
\vspace{0.3cm}
\includegraphics[scale=0.245]{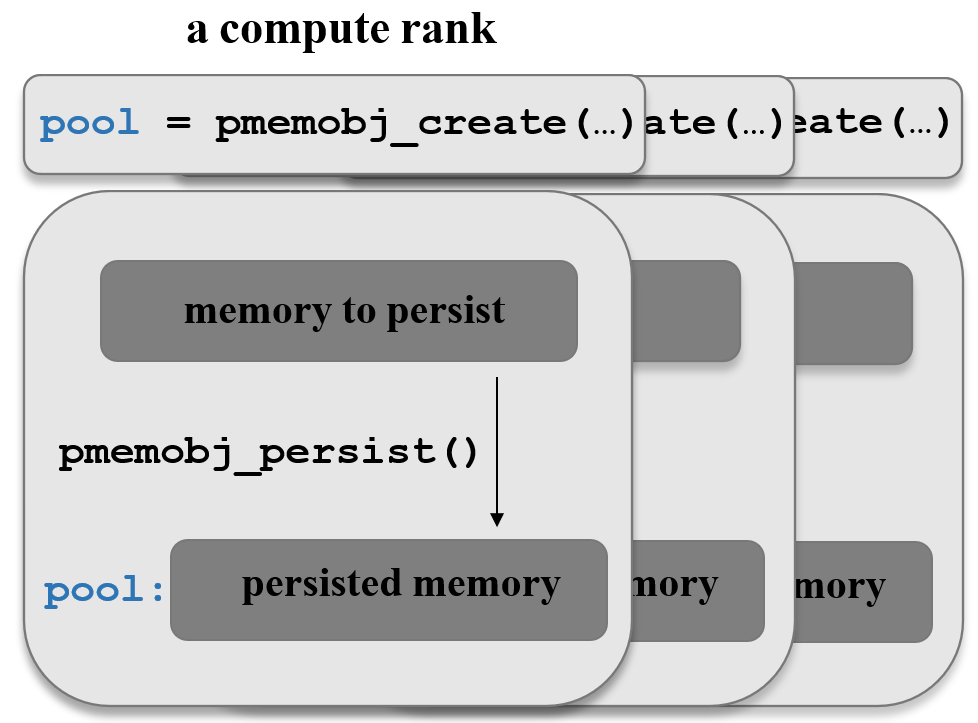}
\caption{Persisting data with \textit{libpmemobj} directly to local NVRAM in the homogeneous NVRAM cluster architecture.}
\label{libpmemobj_in-NVRAM ESR}
\vspace{-0.4cm}
\end{figure}

PDMK~\cite{scargall2020introducing} offers a set of user-space APIs 
and interfaces to interact with non-volatile memory, 
with support for multiple abstraction layers, over Linux and Windows. 
PMDK libraries are designed to leverage the direct
access allowed by persistent memory as much as possible. 
Persistent libraries in PMDK help applications maintain 
data structure consistency in the presence of failures. 
One such library is \textit{libpmemobj}, which helps 
the programmer manage persistent memory arrays and data structures. 
We implement \textit{in-NVRAM} ESR using the \textit{libpmemobj} library 
to persist ESR data directly to local NVRAM in the homogeneous NVRAM cluster architecture. 
Each process first creates a persistent memory pool 
using a call to \texttt{pmemobj\_create}
and then, at each persistence iteration, persists ESR data using \texttt{pmemobj\_persist}. 
Fig.~\ref{libpmemobj_in-NVRAM ESR} illustrates a persistence iteration 
using \textit{libpmemobj} in the homogeneous NVRAM cluster architecture.

PMDK also provides \textit{librpmem}, a remote RDMA access library, 
which supports remote access to persistent memory, 
with a synchronous write model:
The local initiator writes and all of the remotely replicated 
writes must complete before the local write returns to the application. 
This library can be useful in the implementation of \textit{in-NVRAM} ESR 
in the PRD architecture, although this is out of the scope of this paper.

\subsection{Persistent Memory File Systems (PMFSs)}
\label{sec:implementation PMFS}

We also consider persisting data to NVRAM using 
PMFSs, which can be either local or distributed. 
PMFSs often exploit the byte-addressability of NVRAM,
and support a special Direct Access (DAX) mode,
which enables memory mapping directly from the NVRAM to the application memory space. 
DAX bypasses the kernel, page cache, and I/O subsystem, 
avoids interrupts and context switching, 
and allows the application to perform byte-addressable 
\textit{load/store} operations~\cite{intel-quick-start-guide-persistent-memory3}.

Fridman et al.~\cite{fridman2021assessing} present an evaluation of local PMFSs 
(e.g., ext4-DAX and SplitFS~\cite{kadekodi2019splitfs})
for writing and reading diagnostics of scientific applications, 
as well as for performing C/R of these applications using transparent 
or explicit checkpointing. 
While these local PMFSs are good choices for single-node workloads, 
scientific computing applications typically require a distributed 
PMFS operating on multiple nodes. 
\textit{in-NVRAM} ESR can utilize local PMFSs in the homogeneous NVRAM cluster 
architecture to persist the recovery data of each process locally in its node. 
Distributed PMFSs, however, control how data is stored and retrieved 
from NVM when accessed by more than one node, using network communication. 
Distributed PMFSs can be utilized by \textit{in-NVRAM} ESR in the PRD sub-cluster 
architecture to persist recovery data remotely on the NVRAM present on PRD nodes. 

In this work, we implement \textit{in-NVRAM} ESR for the homogeneous NVRAM 
cluster architecture by using ext4-DAX as a local PMFS over the local NVRAM.

\section{Key Example: PCG Solver}\label{key}

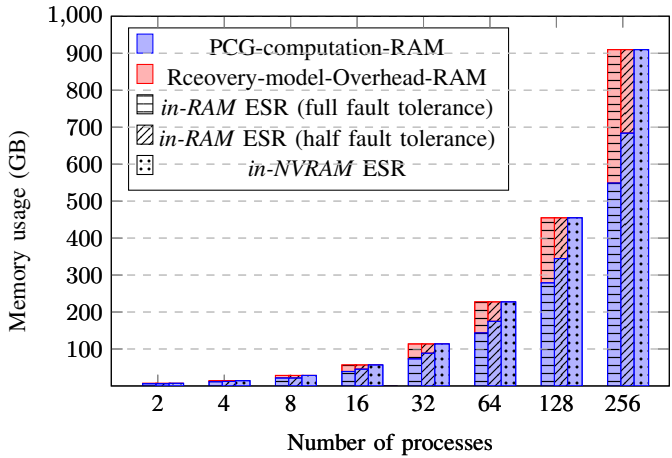
\begin{figure}[ht!]
\small
\begin{subfigure}{0.01\textwidth}
\begin{turn}{90} 
Memory usage (GB)
\end{turn} 
\end{subfigure}%
~
\begin{subfigure}[t!]{0.46\textwidth}
\centering
\begin{tikzpicture}
\pgfplotsset{
every axis/.style={
    xlabel={Number of processes},
    ybar stacked,
    ymin=0, ymax=1000,
    ymajorgrids=true,
    grid style=dashed,
    xtick= {2,4,8,16,32,64,128,256},
    xticklabels={2,4,8,16,32,64,128,256},
    xmode=log,
    legend pos=north west,
    ytick={100,200,300,400,500,600,700,800,900,1000},
    bar width=5.5pt,
    width=9cm,
    height=6.5cm
  },
}

\begin{axis}[hide axis]
\addplot coordinates {(2,0000000)};
\addplot coordinates {(2,0000000)};
\legend{PCG-computation-RAM, Rceovery-model-Overhead-RAM}
\addlegendentry{\textit{in-RAM} ESR (full fault tolerance)}
\addlegendimage{draw=black, pattern=horizontal lines}
\addlegendentry{\textit{in-RAM} ESR (half fault tolerance)}
\addlegendimage{draw=black, pattern=north east lines}
\addlegendentry{\textit{in-NVRAM} ESR}
\addlegendimage{draw=black, pattern=dots}
\end{axis}

\begin{axis}[bar shift=-3pt]
\addplot +[
    postaction={
    pattern=horizontal lines
    }
] coordinates {(2,6.044) (4,11.450) (8, 21.475) (16, 39.150) (32, 73.279) (64, 143.286) (128, 278.928) (256, 548.712)};
\addplot +[
    postaction={
    pattern=horizontal lines
    }
]  coordinates {(2,1.064) (4,2.766) (8,6.955) (16,17.710) (32, 40.451) (64, 84.174) (128, 175.992) (256,361.128)};
\end{axis}

\begin{axis}[bar shift=2pt]
\addplot +[
    postaction={
    pattern=north east lines
    }
] coordinates {(2,6.044) (4,12.200) (8, 21.444) (16, 45.280) (32, 88.430) (64, 174.688) (128, 344.560)(256, 683.800) };
\addplot +[
    postaction={
    pattern=north east lines
    }
] coordinates {(2,1.064) (4,2.016) (8,6.986) (16,11.580) (32, 25.300) (64, 52.772) (128, 110.360) (256, 226.040 )};
\end{axis}
\begin{axis}[bar shift=7pt, hide axis]
\addplot +[
    postaction={
    pattern= dots
    }
] coordinates {(2,7.108) (4,14.216) (8, 28.430) (16, 56.860) (32, 113.730) (64, 227.460) (128, 454.920) (256,909.840)};
\addplot +[
    postaction={
    pattern= dots
    }
] coordinates {(2,0) (4,0) (8,0) (16,0) (32,0) (64,0) (128,0)};
\end{axis}
\end{tikzpicture}
\end{subfigure}%
\vspace{-0.1cm}
\caption{RAM usage for calculation and recoverability PCG for a 7-point stencil of a 3-D Poisson equation.}\label{figure:memory overhead}
\end{figure}

We focus on the \emph{preconditioned conjugate gradient} (\emph{PCG}) 
iterative solver also studied in prior ESR research, 
because it is commonly used for solving sparse linear systems and employed by the representative HPCG scientific benchmark \cite{marjanovic2014performance}. PCG solves the linear equation $Ax=b$ for a symmetric positive definite matrix $A_{n \times n}$. An ESR algorithm for PCG has been derived by the generic meta-algorithm described in~\cite{pachajoa2020generic}. To recover from node failures, PCG saves redundancies for the search direction variable. (see Appendix~\ref{appendix} for more details). 
We use our experimental cluster (which contains 8 nodes), described in Table~\ref{table:cluster_specs}, to simulate the recovery patterns in a larger cluster of 256 nodes, as each core acts as a difference compute node. Hence, to support full fault tolerance ($\phi=\#processes$), each process sends redundancies to all other processes. To support the recovery in more reliable systems, each process sends redundancies to a certain fraction of the processes ($\phi=\alpha\cdot\#processes$). In our experiments we examine half fault tolerance, when recovery data is copied to half of the processes ($\alpha=0.5$).
Fig.~\ref{figure:memory overhead} depicts the memory usage of PCG for the 7-point stencil of a 3-D Poisson equation. RAM utilization is fixed per process, so the total memory usage increases linearly with the number of processes. The figure shows that the size of problems that can be handled using \textit{in-RAM} ESR decreases because recovery data occupy a significant part of the node DRAM.

\begin{table}[ht]
{\scalebox{.8}{
\begin{tabular}{ |c|c|c| } 
 \hline
  & \textbf{Compute Nodes} & \textbf{NVRAM Storage Node} \\ 
  \hline
  \#Nodes & 8 & 1  \\ 
 \hline
 \#Sockets & 2 (per node) & 2\\ 
 \hline
 CPU Spec. & 16 Cores $\times$ Intel(R) & 10 Cores $\times$ Intel(R) \\ & Xeon(R) Gold 6130 CPU &  Xeon(R) Gold 5215 CPU \\ & @ 2.10GHz (per socket) & @ 2.50GHz (per socket) \\ 
 \hline
 L1 Cache & \multicolumn{2}{c|}{32KB i-Cache 32KB d-Cache (per core)} \\ 
 \hline
 L2 Cache & \multicolumn{2}{c|}{1024KB (per core)} \\ 
 \hline
 L3 Cache & 22528KB (shared, per socket) & 14080KB (shared, per socket) \\ 
 \hline
 DRAM Spec. & 32GB DDR4 DRAM 2666 MT/s & 16GB DDR4 DRAM 2933 MT/s \\ 
 \hline
 Total DRAM & 128GB (per node) & 192GB  \\ 
 \hline
 NVM Spec. & none &  256GB Intel Optane™ DCPMM \\ & & 2666 MT/s Apache Pass \\ 
 \hline
 Total NVM & none & 1024GB [(2 sockets)$\times$ \\ & & (2 channels)$\times$256GB] \\ 
 \hline
 SSD Spec. & none & 240GB 6GB/s Intel SATA 2.5” SSD \\ 
 \hline
 Network & \multicolumn{2}{c|}{56Gb/s Mellanox Infiniband FDR} \\
 \hline
 HT & \multicolumn{2}{c|}{disabled} \\ 
 \hline
 OS & Linux CentOS 7 &  Linux CentOS 7.9 \\ 
 \hline
\end{tabular}
}}
\caption{Experimental cluster specifications.}
\label{table:cluster_specs}
\end{table}

\paragraph*{Extrapolation to Aurora}
Aurora~\cite{stevens2019aurora} will 
consist of 9000 compute nodes, each with 112 CPU cores ($\sim$ $10^6$ cores total). 
Total system memory is estimated at 10PB. 
It is expected to have over 230PB of high-performance storage, 
including Intel Optane™ DC SSDs and DCPMMs (with DAOS). 
For a PCG 7-point stencil of a 3-D Poisson equation, 
extrapolating the \textit{in-RAM} ESR PCG RAM consumption presented in Fig.~\ref{figure:memory overhead} 
into Aurora scale, we estimate \textit{in-RAM} full fault tolerance 
ESR RAM consumption to be $\sim$30\% of the system memory, hence $\geq$ 3PB. 
\textit{In-NVRAM} ESR suggests eliminating this memory overhead at the expense 
of only $3$PB $\div 9000 = \sim 0.3$TB usage of NVRAM because every value 
that resides in the RAM of $\sim9000$ nodes can now be persisted to NVRAM only once (or with certain RAID level). 


\subsection{Evaluation}
\label{subsec:evaluation}

\begin{figure}[ht]
\captionbox{DCPMM population configuration of the NVRAM Storage Node, with two sockets (SO) connected via an interconnect. 
Each CPU has two memory controllers (MC), each providing three memory channels (CH); 
each memory channel contains two DIMM slots. 
D (red) denotes DCPMM, R (blue) denotes DRAM (RDIMM); white denotes vacancy. 
\label{dcpmm_population}}
{\scalebox{.48}{
\includegraphics[width=\textwidth]{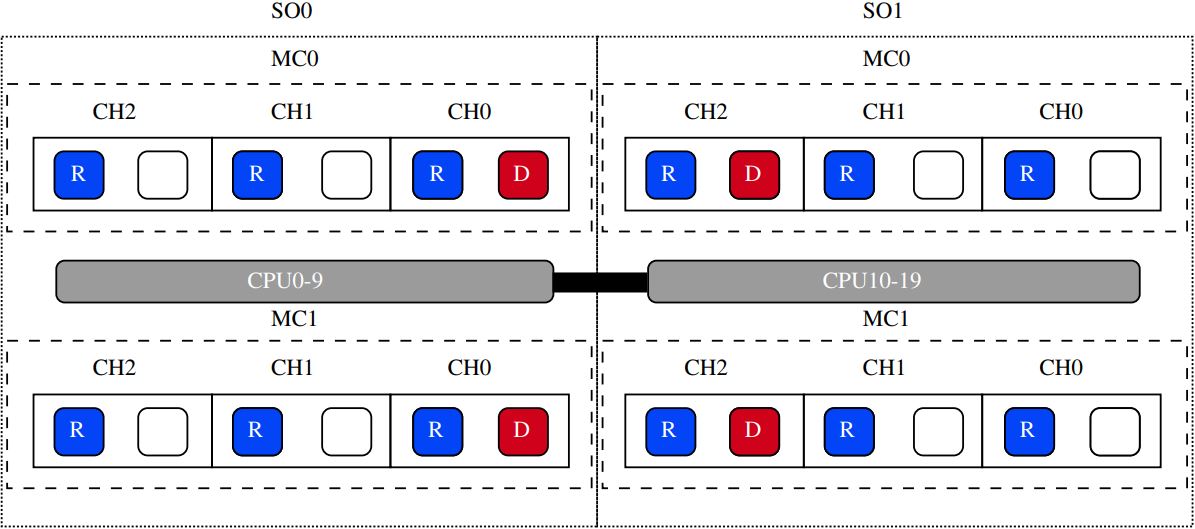}}}
\end{figure}

To evaluate \textit{in-NVRAM} ESR's performance, we employed it for the PCG solver and compared it with the fully fault tolerant ESR. 
Our experimental cluster specifications are listed in Table~\ref{table:cluster_specs}. 
Our cluster consists of 8 compute nodes (each with 32 compute cores and 128GB DRAM) 
and a single NVRAM node (with 20 compute cores, 192GB DRAM, and 1TB Intel Optane™ DCPMM, 
populated as depicted by Fig.~\ref{dcpmm_population}).

We evaluated both the homogeneous NVRAM cluster and 
the NVRAM PRD sub-cluster architectures. 
As our cluster contains only a single NVRAM node, 
our evaluation of the homogeneous NVRAM cluster architecture is limited to 20 processes. 
Nevertheless, this evaluation can be reliably extrapolated to 
a multi-node homogeneous NVRAM cluster architecture, 
since persisting data from each process to its local node is 
an embarrassingly parallel workload at the node level. 

To evaluate \textit{in-NVRAM} ESR/PRD, we use the NVRAM node in 
our experimental cluster as a single PRD node. 
Multiple NVRAM nodes can serve as the PRD sub-cluster, 
distributing the persistent data to eliminate bandwidth bottlenecks 
or creating RAID over the sub-cluster to increase fault tolerance.

\begin{figure}[]
\begin{subfigure}{0.01\textwidth}
\vspace{-0.8cm}
\begin{turn}{90} 
NVRAM Usage (GB)
\end{turn} 
\end{subfigure}%
~
\begin{subfigure}{0.25\textwidth}
\centering
\begin{tikzpicture}
\pgfplotsset{
every axis/.style={
    xlabel={Number of processes},
    ybar stacked,
    ymin=0, ymax=3,
    ymajorgrids=true,
    grid style=dashed,
    xtick= {2,4,8,16,32,64,128,256},
    ticklabel style = {font=\footnotesize},
    xticklabels={2,4,8,16,32,64,128,256},
    xmode=log,
    legend pos=north west,
    ytick={1, 2, 3},
    bar width=7pt,
    width=5.4cm,
    height=4.7cm
  },
}

\begin{axis}[]
\addplot [color=olive, fill] coordinates {(2,0.57) (4,0.89) (8,1.296) (16,1.694) (32,2) (64,2.21) (128,2.3) (256, 2.4)};
\end{axis}
\end{tikzpicture}
\end{subfigure}%
\begin{subfigure}{0.23\textwidth}
\centering
\begin{tikzpicture}
\pgfplotsset{
every axis/.style={
    xlabel={Global Vector Size (x1000)},
    ybar stacked,
    ymin=0, ymax=3,
    ymajorgrids=true,
    grid style=dashed,
    xtick= {5000,10000,20000,40000,80000,160000,320000},
    xticklabels={5,10,20,40,80,160,320},
    ticklabel style = {font=\footnotesize},
    xmode=log,
    legend pos=north west,
    ytick={1, 2, 3},
    bar width=7.5pt,
    width=5.2cm,
    height=4.7cm
  },
}

\begin{axis}[]
\addplot [color=olive, fill] coordinates {(5000,0.04)(10000,0.08)(20000,0.16) (40000,0.32) (80000,0.64) (160000,1.28) (320000,2.56)};
\end{axis}
\end{tikzpicture}
\end{subfigure}%
\vspace{-0.1cm}
\caption{NVRAM usage by \textit{in-NVRAM} ESR as a function of the number of processes in a node (left) and as a function of the problem's vector size (right).}
\label{NVRAM-usage}
\end{figure}
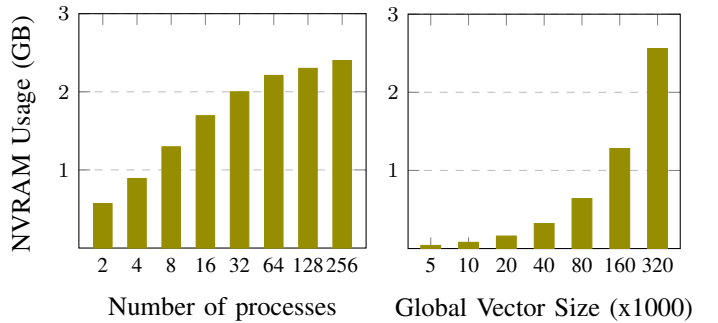

Fig.~\ref{NVRAM-usage} shows an estimate of NVRAM utilization 
of \textit{in-NVRAM} ESR on our cluster. 
The graph on the left shows the amount of NVRAM 
required for different numbers of processes when a fixed amount of RAM is available to each process, and the size of the problem grows to fit available RAM.
The graph on the right shows the NVRAM utilization for different 
sizes of the global input vector of the problem. 
As the global vector is split between the processes 
and each process persists its local parts to the NVRAM, 
the total number of values that are persisted to NVRAM 
is equal to the global vector size.

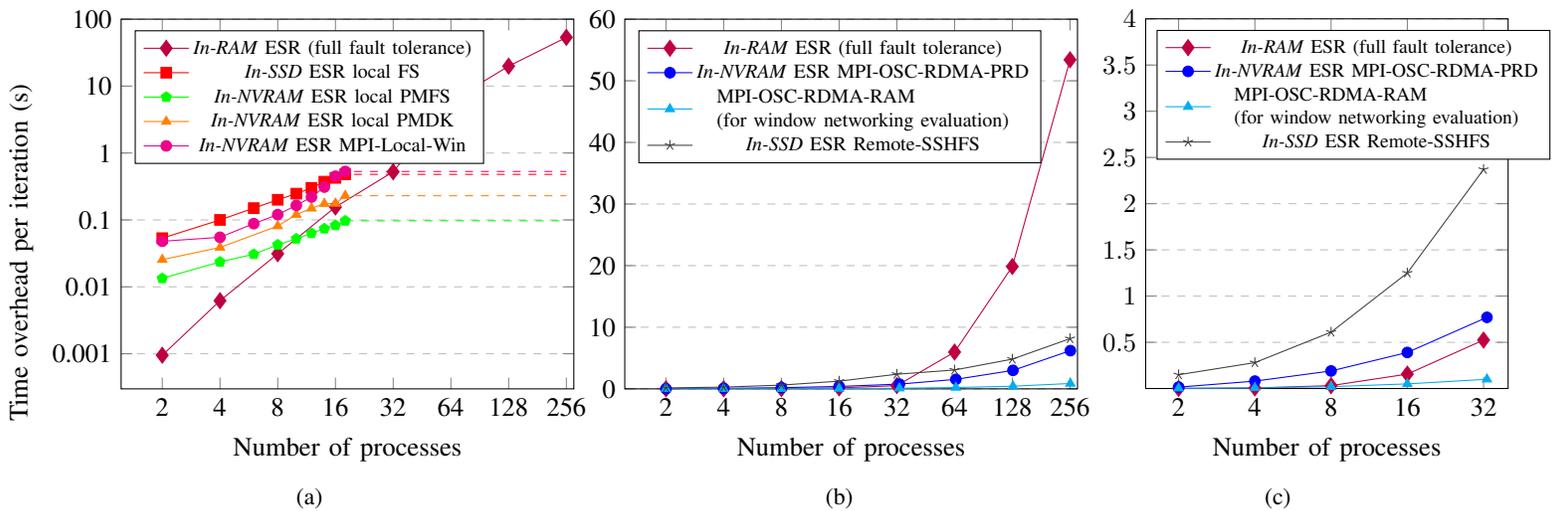
\begin{figure*}
\hspace{-1.3cm}
\begin{subfigure}{0.025\textwidth}
\begin{turn}{90} 
\vspace{-4cm}
Time overhead per iteration (s)
\end{turn} 
\end{subfigure}%
\begin{subfigure}[]{0.39\linewidth}
\centering
\begin{tikzpicture}
\begin{axis}[
    xlabel={Number of processes},
    xmin=0, xmax=280,
    ymin=0, ymax=100,
    xtick= {2,4,8,16,32,64,128,256},
    xticklabels= {2,4,8,16,32,64,128,256},
    ytick={0.001,0.01,0.1,1,10,100},
    yticklabels={0.001,0.01,0.1,1,10,100},
    ymode=log,
    xmode=log,
    log basis y={10},
    legend pos=north west,
    ymajorgrids=true,
    grid style=dashed,
    width=7.6cm,
    height=6.5cm,
    legend style={nodes={scale=0.75, transform shape}}
]
   \addplot[
    color=purple,
    mark=diamond*,
    mark size=3pt,
    ]
    coordinates {
    (2,0.00095)(4,0.0062)(8,0.0312)(16,0.1558)(32,0.5243)(64,5.9711)(128,19.85)(256,53.4076)
    };
    
    \addplot[
    color=red,
    mark=square*,
    ]
    coordinates {
      (2,0.0534)(4,0.10)(6,0.15)(8,0.20)(10,0.247)(12,0.301)(14,0.373)(16,0.425)(18,0.48)
    };
     \addplot[
    color=green,
    mark=pentagon*,
    ]
    coordinates {
     (2,0.0134)(4,0.0236)(6,0.0307)(8,0.0425)(10,0.0523)(12,0.0632)(14,0.074)(16,0.083)(18,0.097)
    };
    \addplot[
    color=orange,
    mark=triangle*,
    ]
    coordinates {
    (2,0.0255)(4,0.0388)(8,0.081)(10,0.12)(12,0.149)(14,0.175)(16,0.176)(18,0.23)
    };
       \addplot[
    color=magenta,
    mark=*,
    ]
    coordinates {
    (2,0.048)(4,0.055)(6,0.088)(8,0.12)(10,0.165)(12,0.22)(14,0.31)(16,0.45)(18,0.53)
    };

\addplot[
    color=red,
   dash pattern=on 3pt ,
    ]
    coordinates {
    (20,0.48)(256,0.48)
    };
       \addplot[
    color=orange,
   dash pattern=on 3pt ,
    ]
    coordinates {
    (20,0.23)(256,0.23)
    };
          \addplot[
    color=green,
   dash pattern=on 3pt ,
    ]
    coordinates {
    (18,0.097)(256,0.097)
    };
           \addplot[
    color=magenta,
   dash pattern=on 3pt ,
    ]
    coordinates {
    (20,0.53)(256,0.53)
    };
    \legend{\textit{In-RAM} ESR (full fault tolerance), \textit{In-SSD} ESR local FS, \textit{In-NVRAM} ESR local PMFS, \textit{In-NVRAM} ESR local PMDK, \textit{In-NVRAM} ESR MPI-Local-Win}

\end{axis}
\end{tikzpicture}
\subcaption{}
\label{homogeneous_time_overhead}
\end{subfigure}%
\hfill
\begin{subfigure}[]{0.38\linewidth}
\centering
\begin{tikzpicture}
\begin{axis}[
    xlabel={Number of processes},
    xmin=0, xmax=280,
    ymin=0, ymax=60,
    xtick= {2,4,8,16,32,64,128,256},
    xticklabels= {2,4,8,16,32,64,128,256},
    xmode=log,
    ytick={0,10,20,30,40,50,60},
    yticklabels={0,10,20,30,40,50,60},
    legend pos=north west,
    ymajorgrids=true,
    grid style=dashed,
    width=7.6cm,
    height=6.5cm,
    legend style={nodes={scale=0.75, transform shape, align=left}}
]

\addplot[
    color=purple,
    mark=diamond*,
    mark size=3pt,
    ]
    coordinates {
    (2,0.00095)(4,0.0062)(8,0.0312)(16,0.1558)(32,0.5243)(64,5.9711)(128,19.85)(256,53.4076)
    };
\addplot[
    color=blue,
    mark=*,
    ]
    coordinates {
    (2,0.015)(4,0.08)(8,0.19)(16,0.39)(33,0.77)(65,1.54)(129,3)(257,6.2)
    };
       
\addplot[
    color=cyan,
    mark=triangle*,
    ]
    coordinates {
      (2,0.003)(4,0.007)(8,0.02)(16,0.05)(33,0.1)(65,0.22)(129,0.43)(257,0.87)
    };
\addplot[
    color=darkgray,
    mark=star,
    ]
    coordinates {
    (2,0.149)(4,0.28)(8,0.61)(16,1.25)(32,2.37)(64,3.06)(128,4.82)(256,8.15)
    };
    \legend{\textit{In-RAM} ESR (full fault tolerance), \textit{In-NVRAM} ESR MPI-OSC-RDMA-PRD, MPI-OSC-RDMA-RAM \\ (for window networking evaluation),
    \textit{In-SSD} ESR Remote-SSHFS
    }
\end{axis}
\end{tikzpicture}
\subcaption{}
\label{PRD_time_overhead}
\end{subfigure}%
\hfill
\hspace{-0.25cm}
\begin{subfigure}[]{0.27\linewidth}
\centering
\begin{tikzpicture}
\begin{axis}[
    xlabel={Number of processes},
    xmin=0, xmax=40,
    ymin=0, ymax=4,
    xtick= {2,4,8,16,32},
    xticklabels= {2,4,8,16,32},
    xmode=log,
    ytick={0.5,1,1.5,2,2.5,3,3.5,4},
    legend pos=north west,
    ymajorgrids=true,
    grid style=dashed,
    width=6.4cm,
    height=6.5cm,
    legend style={nodes={scale=0.73, transform shape, align=left}}
]

\addplot[
    color=purple,
    mark=diamond*,
    mark size=3pt,
    ]
    coordinates {
    (2,0.00095)(4,0.0062)(8,0.0312)(16,0.1558)(32,0.5243)
    };
\addplot[
    color=blue,
    mark=*,
    ]
    coordinates {
    (2,0.015)(4,0.08)(8,0.19)(16,0.39)(33,0.77)
    };
\addplot[
    color=cyan,
    mark=triangle*,
    ]
    coordinates {
      (2,0.003)(4,0.007)(8,0.02)(16,0.05)(33,0.1)
    };
\addplot[
    color=darkgray,
    mark=star,
    ]
    coordinates {
    (2,0.149)(4,0.28)(8,0.61)(16,1.25)(32,2.37)
    };
    \legend{\textit{In-RAM} ESR (full fault tolerance), \textit{In-NVRAM} ESR MPI-OSC-RDMA-PRD, MPI-OSC-RDMA-RAM \\ (for window networking evaluation),
    \textit{In-SSD} ESR Remote-SSHFS,
    }
\end{axis}
\end{tikzpicture}
\subcaption{}
\label{PRD_time_overhead_zoom}
\end{subfigure}
\caption{(a) Time overhead (in log-scale) for \textit{in-NVRAM} ESR and \textit{in-RAM} ESR for single persistence/redundancy iteration in crash-free CG computation in the homogeneous cluster architecture. (b) Time overhead for \textit{in-NVRAM} ESR and \textit{in-RAM} ESR for single persistence/redundancy iteration in crash-free CG computation in the PRD sub-cluster architecture. (c) Zoom in (in log-scale) on  chart \ref{PRD_time_overhead} with $\leq 32$ processes. }
\end{figure*}

We next present the time overheads of a single persistence iteration 
in \textit{in-NVRAM} ESR PCG for a 7-point stencil of a 3-D Poisson equation 
(with a fixed size for local vectors of 176,400 entries each). 
The time overheads of a single redundancy iteration of \textit{in-RAM} 
ESR with full fault tolerance are presented as well. 
We focus on the time overhead of a single persistence iteration. 
We do not show time overheads for the reconstruction phase because the number of reconstruction phases throughout the execution 
can be assumed to be much smaller than that of persistence iterations. 
Moreover, the reconstruction phase is heavily governed by the full-state 
reconstruction calculations, as explained in~\cite{ChenHPDC2011}. 

Fig.~\ref{homogeneous_time_overhead} presents the time overheads of ESR and \textit{in-NVRAM} ESR 
in the homogeneous cluster architecture. We have implemented persistence to the local NVRAM using the ext4-dax local PMFS, PMDK, and MPI local windows over NVRAM. 
For reference, we also measured the time overhead of persisting 
the ESR data to a local SATA-SSD device. 
Dashed lines refer to a natural extrapolation of the results beyond 
a single NVRAM node\footnote{We remind the reader that our single NVRAM node contains 20 cores.}, based on the simple observation that local persistence operations in different nodes proceed in parallel. Above 32 processes, ESR's time overhead increases significantly since persistence data must be sent to the RAM of processes in remote nodes. 

Fig.~\ref{PRD_time_overhead} presents the evaluation of a single 
persistence iteration of \textit{in-NVRAM} ESR in the PRD sub-cluster architecture 
implemented with MPI OSC over RDMA to NVRAM. 
To show the cost of implementing MPI OSC over NVRAM,
we also present the time overhead of MPI OSC over RDMA when the windows are on \emph{RAM} without persistent operations. 
For reference, we measure the time overhead of persisting the ESR data to a remote SATA-SSD device via SSH-FS.
It can be seen that the overhead of ensuring persistence 
is relatively small.
It can also be seen that MPI-OSC over RDMA access to NVRAM 
is faster than accessing a remote SSD storage device, 
especially at high process counts. 

These results show that \textit{in-NVRAM} ESR/PRD is significantly superior to ESR in terms of the time overhead incurred by writing the recovery data, 
except when the number of processes is small, and they all fit inside a single node. 

\paragraph*{Acknowledgment}
This work was supported by Pazy grant 226/20, ISF grant 1425/22, and 
the Lynn and William Frankel Center for Computer Science.
Computational support was provided by the NegevHPC project~\cite{negevhpc}. 
The authors would like to thank Gabi Dadush, Israel Hen and Emil Malka 
for their hardware support.

\bibliographystyle{IEEEtran}
\bibliography{IEEEabrv,references.bib}
\clearpage
\appendices
\section{\textit{In-NVRAM} ESR for PCG Solver}
\label{appendix}
\begin{figure}[H]
\centering
 \includegraphics[width=8.5cm,height=3.5cm]{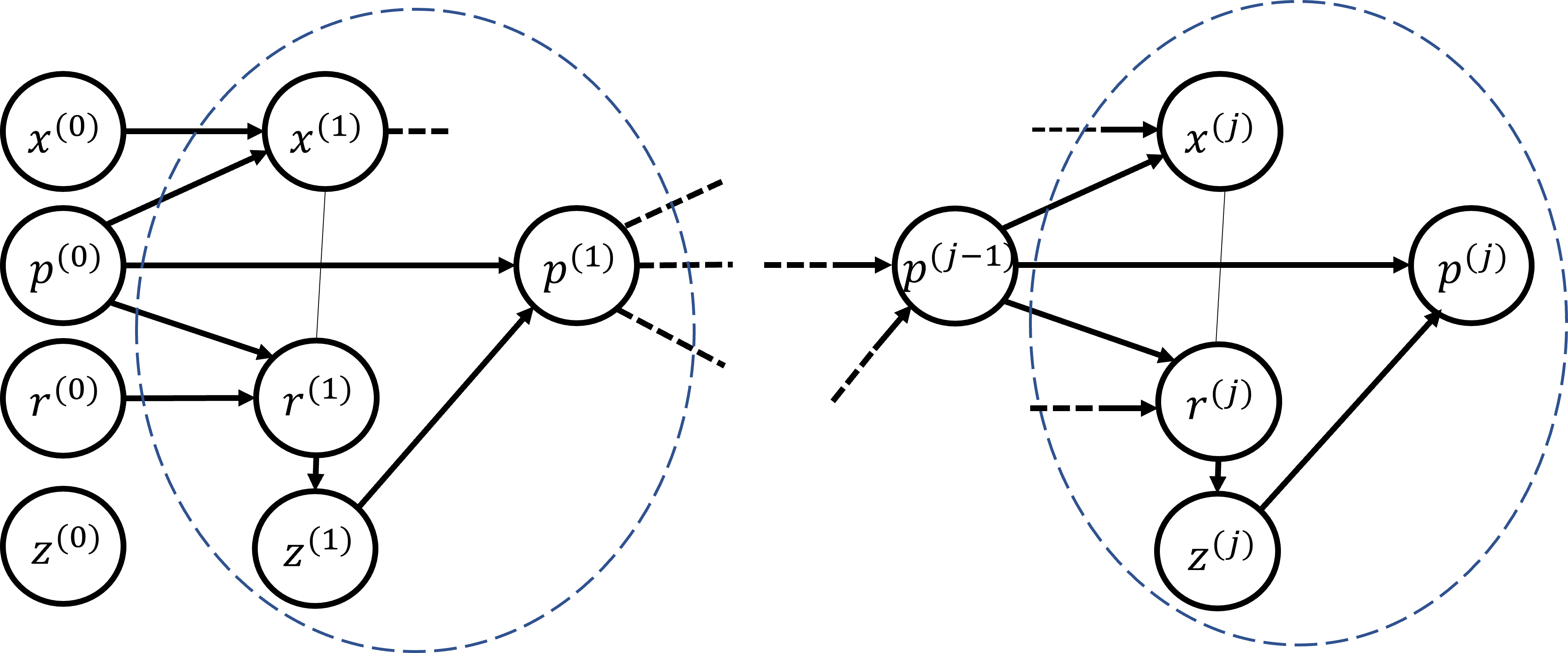}
 \caption{PCG Dependency Graph.}
  \label{fig:pcg_dg}
\end{figure}

Preconditioned Conjugate Gradient (PCG) is a main example of linear iterative solver, as it is commonly used to solve linear systems for a wide range of applications including computational fluid dynamics, wind energy, and particle physics~\cite{eller2019using}. PCG solves the linear equation $Ax=b$ for a symmetric positive definite matrix $A_{n \times n}$
(see Algorithm~\ref{alg:pcg}). The dependency graph of PCG is shown in Fig.~\ref{fig:pcg_dg}. An ESR algorithm for PCG has been derived by the generic meta-algorithm described in~\cite{pachajoa2020generic}, from PCG's dependency graph. To recover from node failures, PCG saves redundancies for the search direction variable, i.e., the variable $p$ (hence $V_{SpMV} = \{p\}$). To reconstruct the complete state of a failed process $f$, two successive values of $p_{I_f}$ are required. 
Therefore, the redundancy of $p$ is saved for two successive iterations (hence $k=2$). Algorithm ~\ref{alg:esr_pcg} describes a persistence iteration of \textit{in-RAM} ESR for PCG (Algorithm~\ref{alg:pcg}), where SpMV operations with $p$ are augmented to ASpMV operations to create $p$'s redundancies. Algorithm~\ref{alg:esr_pcg_reconstruction} shows the reconstruction stage of \textit{in-RAM} ESR, where redundancies of $p$ are collected from the RAM of the surviving nodes to reconstruct the failed processes.

With \textit{in-NVRAM} ESR the redundancies of $p$ are saved for each process in the NVRAM. Algorithm~\ref{alg:nvm-esr_pcg} shows a persistence iteration of PCG, in which values of $p$ are saved in NVRAM (locally or remotely). Algorithm~\ref{alg:nvm-esr_pcg_reconstruction} shows the reconstruction stage of PCG under \textit{in-NVRAM} ESR, collecting $p$'s redundancies by accessing the NVRAM.

\begin{algorithm}[H]
\caption{PCG solver for $Ax=b$.}
\label{alg:pcg}
\begin{algorithmic}[1]
\State $r^{(0)} \gets b-Ax^{(0)}$ , $z^{(0)} \gets Pr^{(0)}$ , $p^{(0)} \gets z^{(0)}$  
\For{$j = 0,1,...$ until convergence}
    \State $\alpha^{(j)} \gets r^{(j)T}z^{(j)}/r^{(j)T}Ap^{(j)}$
    \State $x^{(j+1)} \gets x^{(j)} + \alpha^{(j)}p^{(j)}$
    \State $r^{(j+1)} \gets r^{(j)} - \alpha^{(j)}Ap^{(j)}$
    \State $z^{(j+1)} \gets Pr^{(j+1)}$
    \State $\beta^{(j)} \gets r^{(j+1)T}z^{(j+1)}/r^{(j)T}z^{(j)}$
    \State $p^{(j+1)} \gets z^{(j+1)} + \beta^{(j)}p^{(j)}$

\EndFor
\end{algorithmic}
\end{algorithm}

\begin{algorithm}[H]
\caption{\textit{in-RAM} ESR for each iteration $j$ of two successive redundancy iterations of PCG. Line numbering refers to the lines of Algorithm \ref{alg:pcg}.}
\label{alg:esr_pcg}
\begin{algorithmic}[1]
\makeatletter
\setcounter{ALG@line}{2}
\makeatother
\State $\alpha^{(j)} \gets r^{(j)T}z^{(j)}/r^{(j)T}ASpMV(A,p^{(j)})$
\Statex \vspace{-0.1cm} \hspace{-0.3cm}\vdots
\setcounter{ALG@line}{7}
\State $p^{(j+1)} \gets z^{(j+1)} + \beta^{(j)}p^{(j)}$
\end{algorithmic}
\end{algorithm}

\begin{algorithm}[H]
\caption{\emph{In-RAM} ESR Reconstruction phase of PCG.}
\label{alg:esr_pcg_reconstruction}
\begin{algorithmic}[1]
\State Retrieve the static data $A_{I_f,I},~ P_{I_f,I}~$ and $b_{I_f}$  
\State Gather $r_{I\setminus I_f}^{(j)}$ and $x_{I\setminus I_f}^{(j)}$ 
\State Retrieve the redundant copies of $\beta^{(j-1)}, ~ p_{I_f}^{(j-1)}$ and $p_{I_f}^{(j)}$ from the RAM of other processes
\State Compute $z_{I_f}^{(j)} \gets p_{I_f}^{(j)} - \beta^{(j-1)} p_{I_f}^{(j-1)}  $
\State Compute $v \gets z_{I_f}^{(j)} - P_{I_f,I\setminus I_f}r_{I\setminus I_f}^{(j)} $
\State Solve $P_{I_f,I_f}r_{I_f}^{(j)}=v $ for $r_{I_f}^{(j)}$ 
\State Compute $w \gets b_{I_f}-r_{I_f}^{(j)}-A_{I_f,I\setminus I_f}x_{I\setminus I_f}^{(j)}$
\State Solve $A_{I_f,I_f}r_{I_f}^{(j)}=w $ for $x_{I_f}^{(j)}$
\end{algorithmic}
\end{algorithm}

\begin{algorithm}[H]
\caption{\emph{In-NVRAM} ESR for a persistence iteration $j$ of PCG. Line numbering refers to the lines of Algorithm \ref{alg:pcg}.}
\label{alg:nvm-esr_pcg}
\begin{algorithmic}[1]
\setcounter{ALG@line}{2}
\State $\alpha^{(j)} \gets r^{(j)T}z^{(j)}/r^{(j)T}Ap^{(j)}$
\Statex \vspace{-0.1cm} \hspace{-0.3cm}\vdots
\setcounter{ALG@line}{7}
\State $p^{(j+1)} \gets z^{(j+1)} + \beta^{(j)}p^{(j)}$
\Statex \textbf{if} \textsc{homogeneous Cluster} \textbf{then}
\Statex \ \ \ \  persist $p^{(j+1)}$ to \textbf{local} NVM
\Statex \textbf{if} \textsc{PRD sub-cluster} \textbf{then}
\Statex \ \ \ \  persist $p^{(j+1)}$ to \textbf{remote} NVM
\end{algorithmic}
\end{algorithm}

\begin{algorithm}[H]
\caption{\emph{In-NVRAM} ESR Reconstruction phase of PCG.  Line numbering refers to the lines of Algorithm \ref{alg:esr_pcg_reconstruction}.}
\label{alg:nvm-esr_pcg_reconstruction}
\begin{algorithmic}[1]
\Statex \hspace{-0.6cm} \textbf{if} \textsc{homogeneous Cluster} \textbf{then} 
\Statex \hspace{-0.6cm} \ \ \ \ Wait for failed nodes to recover and have access to \textbf{local} NVM
\Statex \hspace{-0.6cm} \textbf{if} \textsc{PRD sub-cluster} \textbf{then}
\Statex \hspace{-0.6cm} \ \ \ \ Run reconstruction from any spare nodes that have access to the \Statex \hspace{-0.3cm} \textbf{remote} NVRAM Sub-Cluster
\Statex \hspace{-0.6cm} \hrulefill
\Statex Retrieve the static data $A_{I_f,I},~ P_{I_f,I}~$ and $b_{I_f}$  
\Statex Gather $r_{I\setminus I_f}^{(j)}$ and $x_{I\setminus I_f}^{(j)}$ 
\Statex \textbf{if} \textsc{homogeneous Cluster} \textbf{then}
\setcounter{ALG@line}{2}
\State \ \ \ \ Read $\beta^{(j-1)}, ~ p_{I_f}^{(j-1)}$ and $p_{I_f}^{(j)}$ from \textbf{local} NVM
\Statex \textbf{if} \textsc{PRD sub-cluster} \textbf{then}
\setcounter{ALG@line}{2}
\State \ \ \ \ Read $\beta^{(j-1)}, ~ p_{I_f}^{(j-1)}$ and $p_{I_f}^{(j)}$ from \textbf{remote} NVM
\Statex \vspace{-0.2cm} \hspace{-0.3cm} $\vdots$ 
\setcounter{ALG@line}{7}
\State Solve $A_{I_f,I_f}r_{I_f}^{(j)}=w $ for $x_{I_f}^{(j)}$
\end{algorithmic}
\end{algorithm}

\end{document}